\documentclass[11pt]{article}

\usepackage{amsmath,graphicx,amssymb,subfigure}
\usepackage[all]{xy}

\newcommand{\be}{\begin{equation}}
\newcommand{\ee}{\end{equation}}
\newcommand{\beqa}{\begin{eqnarray}}
\newcommand{\eeqa}{\end{eqnarray}}

\def\dd{\mathrm{d}}
\def\hq{\rho}

\def\mn{{\mu\nu}}

\begin{document}

\baselineskip=15.5pt
\thispagestyle{empty}
\setcounter{page}{1}

\newfont{\namefont}{cmr10}
\newfont{\addfont}{cmti7 scaled 1440}
\newfont{\boldmathfont}{cmbx10}
\newfont{\headfontb}{cmbx10 scaled 1728}
\renewcommand{\theequation}{{\rm\thesection.\arabic{equation}}}
\font\cmss=cmss10 \font\cmsss=cmss10 at 7pt
\par\hfill ITF-UU-11/19
\par\hfill SPIN-11/14

\begin{center}
{\LARGE{\bf Black holes and black branes in Lifshitz spacetimes}}
\end{center}
\vskip 10pt
\begin{center}
{\large 
Javier Tarr\'\i o, Stefan Vandoren}
\end{center}
\vskip 10pt
\begin{center}
\textit{Institute for Theoretical Physics and Spinoza Institute, Universiteit Utrecht, 3584 CE, Utrecht, The Netherlands.}

{\small l.j.tarriobarreiro@uu.nl, S.J.G.Vandoren@uu.nl}
\end{center}

\vspace{15pt}

\begin{center}
\textbf{Abstract}
\end{center}

\vspace{4pt}{\noindent 
We construct analytic solutions describing black holes and black branes in asymptotically Lifshitz spacetimes with arbitrary dynamical exponent $z$ and for arbitrary number of dimensions. The model considered consists of Einstein gravity with negative cosmological constant, a scalar, and $N$ $U(1)$ gauge fields with dilatonic-like couplings. We study the phase diagrams and thermodynamic instabilities of the solution, and find qualitative differences between the cases with $1\leq z<2$, $z=2$ and $z>2$.
}
\vfill

\newpage

\section{Introduction and summary of results}

One of the directions in  which the AdS/CFT correspondence \cite{Maldacena:1997re} has been extended in recent years is towards the construction of gravity models conjectured to be dual condensed matter systems with anisotropic scaling \cite{Son:2008ye,Balasubramanian:2008dm,Kachru:2008yh}. It is expected that this direction of research, if successful in finding a dictionary translating between the gravitational degrees of freedom and field theory operators, would shed light onto the non-perturbative dynamics of non-relativistic models with this kind of scaling.

In this work we will be interested in the development of the gravitational dual description of models exhibiting anisotropic scale invariance  of the type
\be
t \to \lambda^z t \, \qquad \vec x \to \lambda \, \vec x \ ,
\ee
where $z$ is called the dynamic exponent. For $z=1$, the scaling is isotropic; it corresponds to relativistic invariance. For generic values of $z$, the system is said to have Lifshitz scaling. The dual boundary field theory is not relativistic, but still allows for particle production. For the special case of $z=2$ there is an extension of the Lifshitz scaling symmetry group to the Schr\"odinger group, in which particle number and special conformal transformations are conserved. See e.g. \cite{Son:2008ye,Balasubramanian:2008dm,Maldacena:2008wh,Ross:2009ar,Balasubramanian:2010uw,Guica:2010sw} for some more references on this topic.

In \cite{Kachru:2008yh} it was proposed that gravity duals of field theories with Lifshitz scaling should have metric solutions that asymptote the form
\be\label{eq.lifshitzmetric}
\dd s^2 = \frac{\ell^2}{r^2} \dd r^2 - \frac{r^{2z}}{\ell^{2z}} \dd t^2 + \frac{r^2}{\ell^2} \dd \vec x_{d-1}^2 \ ,
\ee
which is the generalization of anti-de Sitter spacetime ($z=1$) to non-trivial dynamic exponent. A metric that locally looks like \eqref{eq.lifshitzmetric}, we call a Lifshitz metric; it is invariant under the Lifshitz scaling if we let the radial coordinate scale as $r \to \lambda^{-1} r$. The real parameter $\ell$ represents the radius of AdS when $z=1$, and we will refer to it as the radius of Lifshitz spacetime. 

In a holographic fashion, the dual theory would be formulated in a $d$-dimensional hypersurface located at infinite radial distance, the boundary. A boundary system on $\mathbb{R}_t \times S^{d-1}$ or $\mathbb{R}_t \times \mathbb{R}^d$ with finite temperature and chemical potential can be realized from charged black holes or black branes respectively. In this paper,  we will study the properties of asymptotically Lifshitz black holes and black branes by their own, and will not pay much attention to the holographic connections to non-relativistic field theory. We expect to address this issue in the future, and will have this application in mind throughout the paper.

There is a large number of papers available in the literature discussing black holes in asymptotically Lifshitz spacetimes,  mostly using numerical methods. Some examples of analytic solutions, not necessarily with an Einstein gravity action, exist for fixed value of $z$ \cite{Brynjolfsson:2009ct,Balasubramanian:2009rx,AyonBeato:2009nh,Cai:2009ac,Pang:2009pd,AyonBeato:2010tm,Dehghani:2011tx,Chemissany:2011mb,Maeda:2011jj}. Numerical methods can be also employed to study a continuous range of $z$ for both black holes and black branes, as done for example in \cite{Danielsson:2009gi,Mann:2009yx,Bertoldi:2009vn,Bertoldi:2009dt,Dehghani:2010kd,Brenna:2011gp,Amado:2011nd}.

Most of the papers cited in the previous paragraph are based on an action containing a massive gauge field. In \cite{Taylor:2008tg}, an Einstein-Maxwell-scalar action with $U(1)$ gauge invariance was considered, thereby providing an alternative realization of a gravity model supporting Lifshitz geometries. Both formulations have their own virtues, but also their disadvantages. The formulation with massive gauge fields can be embedded in supergravity models and string compactifications \cite{Maldacena:2008wh,Balasubramanian:2010uk,Donos:2010tu,Cassani:2011sv,Halmagyi:2011xh,Chemissany:2011mb,Amado:2011nd}, but as mentioned above, the study of black holes is mostly numerical or for special values of the dynamical exponent $z$. This makes it less suitable or practical for applications in holography. On the other hand, the background in \cite{Taylor:2008tg}, including a dilaton-like scalar,  is under better analytic control, and analytic charged black hole solutions can be found easily, as we demonstrate in this paper. The 
disadvantage of this model is that the dilaton is not constant and diverges on the boundary, which might be problematic for holography. It also means that the boundary theory is not exactly Lifshitz, but obeys some generalized scaling behavior\footnote{We thank Marika Taylor and Kostas Skenderis for a discussion on this issue.}. Presumably, a proper embedding of this model into string theory will shed more light on this issue.

In this paper we will give analytic solution for black holes and black branes in any number of spacetime dimensions $d\geq3$. This solution will have $z\geq1$ as a continuous parameter and, in a sense, can be seen as a Reissner-Nordstr\"om version of asymptotically Lifshitz black holes. The matter content of the system we study, which is described by the action given later in equation \eqref{eq.action}, consists of $N\geq1$ abelian $U(1)$ fields and one real scalar. It is an extension of the model considered in \cite{Taylor:2008tg}, and is similar to the models studied in \cite{Goldstein:2009cv,Charmousis:2010zz}.  Some of the gauge fields (how many depends on the topology of the black hole studied, as we will see below) and the scalar field are needed to support the Lifshitz spacetime at the boundary, and the remanent matter fields will contribute to Reissner-Nordstr\"om-like terms in the metric, which will have a clear signature in the thermodynamics of the system as a chemical potential term (in the grand-canonical ensemble).

Before summarizing our results in the rest of this section, we find convenient to remind the reader of previous results found in asymptotically AdS ($z=1$) spaces, both for the sake of comparing the solution to known cases and as a basis for our analysis.

In \cite{Hawking:1982dh} it was shown that, for black holes with spherical topology in asymptotically AdS spacetimes, there is a phase transition at a given temperature from a description in terms of thermal AdS (lower temperature) to a black hole setup. This Hawking-Page transition is due to a competing effect between the scale set by the volume of the spacetime and the scale set by the temperature. In \cite{Witten:1998zw} it was generalized to arbitrary dimension, and an explanation in terms of a confinement/deconfinement transition via a dual holographic field theory was given. The existence of unstable small black holes and the corresponding Hawking-Page transition, for the specific value $z=2$, was predicted in \cite{Amado:2011nd} in a setup derived from string theory. We will find that the transition is present in the range $1\leq z\leq2$ in our model.

A $U(1)$ gauge field was included in the setup in \cite{Chamblin:1999tk}, and the properties of charged black holes described by an Einstein-Mawell action with negative cosmological constant were computed. These solutions describe an asymptotically AdS Reissner-Nordstr\"om  black hole with a  horizon topology $ S^{d-1}$ at fixed time. The existence of charged solutions gives rise to a rich phase structure both in the canonical and grand-canonical ensembles, which we  describe below.

The scale determined by the volume of the $S^{d-1}$ plays a crucial r\^ole in the construction of the phase diagrams, as in the case of the Hawking-Page transition commented above. A black brane solution (with a horizon with fixed-time topology $\mathbb{R}^{d-1}$) can be obtained by considering an \emph{infinite volume limit}, which we will denote $\eta\to\infty$, where $\eta$ is a dimensionless parameter to be introduced later. In this case the phase structure becomes trivial, with thermodynamics dominated by black holes for all temperatures.

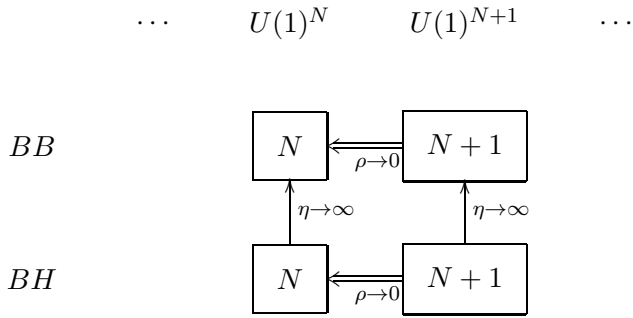
\begin{figure}[ht]
\[
\xymatrix{
 & \cdots  &   U(1)^{N}   &   U(1)^{N+1}  & \cdots \\
 BB &  & *+++[F]{N}  & *+++[F]{N+1} \ar@{=>}[l]^{\hq\to0} & \\
   BH &   & *+++[F]{N} \ar[u]_{\eta\to\infty} & *+++[F]{N+1} \ar[u]_{\eta\to\infty}  \ar@{=>}[l]^{\hq\to0} & 
   }
\]
\caption{Diagram showing the relations between theories with one gauge field less, with boundary topology $\mathbb{R}_t\times S^{d-1}$ (BH, after black hole) and $\mathbb{R}_t\times \mathbb{R}^{d-1}$ (BB, after black brane). The square represents asymptotically AdS spaces. The diagram continues indefinitely to the right and to the left up to $U(1)^0$, consisting on the Schwarzschild-AdS solution. The framed quantities indicate how many charges are there in the black hole.\label{fig.adsdiagram}}
\end{figure}

The results in  \cite{Chamblin:1999tk} can be generalized to consider the abelian group  $U(1)^N$ without a significant change in the phase structure. Of course, the theories with $M<N$ gauge fields can be recovered by setting $N-M$ charge densities, $\hq_i$, to zero. In figure \ref{fig.adsdiagram} we outline a diagram showing the relations between  theories with different number of gauge fields and horizon topology.

\subsection{Summary of results}

In this paper we study the thermodynamic properties of charged asymptotically Lifshitz  black holes. The solution we present is analytic for any value of the dynamic exponent $z\geq1$ and for any number of spacetime dimensions $d>2$. The model we consider is given by
\be\label{eq.action}
S= - \frac{1}{16\pi G_{d+1}} \int \dd^{d+1}x\, \sqrt{-g} \left[  R -2 \Lambda - \frac{1}{2} \left( \partial \phi \right)^2 - \frac{1}{4} \sum_{i=1}^N e^{\lambda_i \phi} F_i^2 \right] \ ,
\ee
with electric fields in the radial direction turned on. Notice that this is a diffeomorphism-invariant action and we look for a metric solution that asymptotically approaches \eqref{eq.lifshitzmetric} that breaks this symmetry explicitely. In order to accommodate a Lifshitz spacetime in Einstein gravity, the presence of extra matter fields is required. The case with $N=1$ was studied for the first time in \cite{Taylor:2008tg}, where a BB solution was found. Strikingly, the solution for the metric in this case does not present the usual Reissner-Nordstr\"om properties that one finds in asymptotically flat/AdS solutions of the Einstein+Maxwell action. In concrete, the blackening function $b(r)$ defined in \eqref{eq.metricbh} (defining the position of the horizon $r_h$ by means of $b(r_h)=0$) has just one non-negative root, and extremal black holes with finite entropy cannot be constructed.

The gauge field is completely determined by the other fields present in the theory. In principle one would expect to have one free parameter associated to the gauge field, given by the constant of motion associated to $A$, since it appears in the action only through its derivatives. However, the requirement of having an asymptotically Lifshitz manifold (\emph{i.e.}, for $z\neq1$) forces a relationship between this constant of motion and the magnitude of the scalar field $\phi$. In a way, the r\^ole of the gauge field is to provide the appropriate potential to support an (asymptotically) Lifshitz spacetime, and the charge associated to the gauge field translates in the asymptotic properties of the manifold, and not in the horizon structure of the black hole.

\begin{figure}[ht]
\[
\xymatrix{
 \cdots   & U(1)^{N-1}   &  U(1)^{N}   &   U(1)^{N+1}  & \cdots \\ 
  & *+++[F]{N-1} & *+++[F]{\,\,\,N\,\,\,} \ar@{=>}[l]_{\hq\to0} & &   \\ 
 &  & *+++[o][F]{N-1} \ar@{-->}[ul]^{z\to1}  & *+++[o][F]{\,\,\,N\,\,\,} \ar@{=>}[l]^{\hq\to0} \ar@{-->}[ul]_{z\to1} & 
  }
\]
\caption{Diagram showing the relations between theories with one  gauge field less and the AdS limit for black brane solutions of our model. Circles represent asymptotically Lifshitz spacetimes whereas squares correspond to asymptotically AdS ones. The diagram continues indifenitely to the right and to the left up to $U(1)^0$. The framed quantities indicate how many charges contribute to give the Reissner-Nordstr\"om-like factor in $b(r)$.\label{fig.bbdiagram}}
\end{figure}
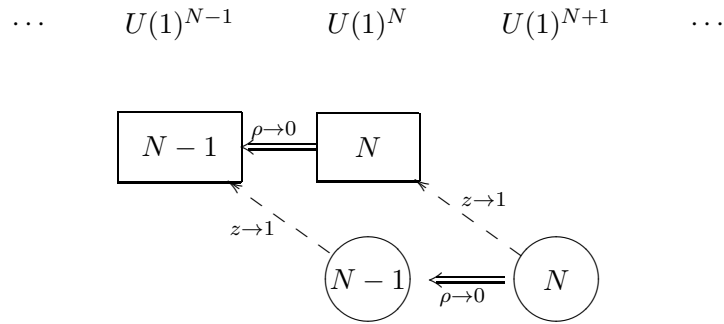

In the next section we present a generalization of this construction with an arbitrary number of $U(1)$ fields. Remarkably, in the black brane case all the $A_{N\geq2}$ gauge fields contribute to give a form of $b(r)$ resembling that of the Reissner-Nordstr\"om solution. Indeed, the constants of motion associated to these extra gauge fields remain undetermined and are interpreted as charge densities $\rho_i$. In general, there is more than one root of $b(r)$ and extremal solutions with a finite horizon size do exist. As in the asymptotically AdS case, we can relate theories with a different number of $U(1)$ fields by consistently turning off charges.  In figure \ref{fig.bbdiagram} we present a sketch of these relations.

Consider now the case with $N=2$ gauge fields. As we just commented, it is possible to find a black brane solution with a rich horizon structure. One would be interested to know whether it is possible to find the correspondent solution but with a black hole, \emph{i.e.}, considering spherical symmetry in the form of the metric\footnote{We are interested in this case and not in the hyperspherical one because it will describe finite size effects in an eventual dual field theory.}. As we will show explicitely in the next section, such construction leads to an algebraic equation whose solution fixes completely the form of both gauge fields in terms of the amplitude of the scalar field. The $b$ function has a form reminiscent of that of the  Schwarzschild-AdS  case. This situation is completely analogous to the one found in \cite{Taylor:2008tg} for the BB solution. In fact, the second gauge field also diverges at the boundary. Therefore, the r\^ole of this gauge field is to support the ``sphericity'' of the solution, and not to modify the horizon structure of the black hole.

\begin{figure}[ht]
\[
\xymatrix{
 \cdots   & U(1)^{N-2}  & U(1)^{N-1}   &  U(1)^{N}   &   U(1)^{N+1}  & \cdots \\ 
& *+++[F]{N-2} & *+++[F]{N-1} \ar@{=>}[l]_{\hq\to0} & & &   \\ 
 & &  & *+++[o][F]{N-2} \ar@{-->}[ull]^{z\to1}  & *+++[o][F]{N-1} \ar@{=>}[l]^{\hq\to0} \ar@{-->}[ull]_{z\to1} & 
  }
\]
\caption{Diagram showing the relations between theories with one  gauge field less and the AdS limit for black hole solutions. Circles represent asymptotically Lifshitz spacetimes whereas squares correspond to asymptotically AdS ones. The diagram continues indifenitely to the right and to the left up to $U(1)^0$. The framed quantities correspond to the number of charges contributing to give the RN-like factor of $b$ .\label{fig.bhdiagram}}
\end{figure}
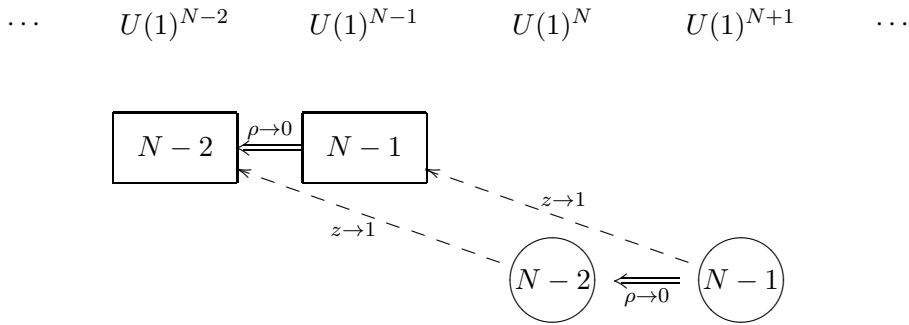

For the spherically symmetric case with more than $2$ $U(1)$ fields, the extra $A_{N\geq3}$ will again contribute in $b$ to a new term resembling the charge-term in the Reissner-Nordstr\"om solution, thus modifying the horizon structure. As we did above, we present in figure \ref{fig.bhdiagram} the relations between theories with different number of gauge fields when we turn off charges or take the AdS limit.

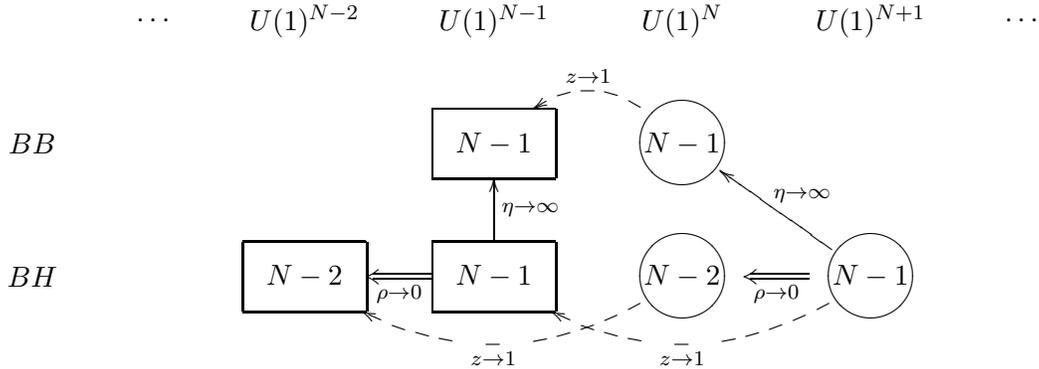
\begin{figure}[h!t]
\[
\xymatrix{
 &  \cdots & U(1)^{N-2}  & U(1)^{N-1}   &  U(1)^{N}   &   U(1)^{N+1}  & \cdots \\
 BB  &  & & *+++[F]{N-1} & *+++[o][F]{N-1} \ar@/_0.7cm/@{-->}[l]_{z\to1} & &   \\
  BH &  &  *+++[F]{N-2}   & *+++[F]{N-1}   \ar@{=>}[l]^{\hq\to0}  \ar[u]_{\eta\to\infty}  & *+++[o][F]{N-2} \ar@/^0.9cm/@{-->}[ll]^{z\to1}  & *+++[o][F]{N-1} \ar[ul]_{\eta\to\infty}  \ar@{=>}[l]^{\hq\to0} \ar@/^0.9cm/@{-->}[ll]^{z\to1} & 
  }
\]
\caption{Relations between the different specific cases arising from action \eqref{eq.action}. See previous captions for symbolism. The web of relations for the cases showed is incomplete for the sake of clarity, but can be completed with the previous diagrams.\label{fig.bbbhdiagram}}
\end{figure}

We have seen that the system described by the action \eqref{eq.action} has a rich web of limiting cases. The question of whether we can recover black brane solutions by taking the appropriate $\eta\to\infty$ limit in the black hole solutions for asymptotically Lifshitz spacetimes has a positive answer. It turns out that this limit effectively decreases the number of $U(1)$ fields in the theory by one, since now the inclusion of a gauge field supporting the ``sphericity'' of the solution is not required anymore. In figure \ref{fig.bbbhdiagram} we present a partial diagram of the relations existing between the black hole and black brane solutions.

\subsubsection{Phase diagrams}

\begin{figure}[tb]
\begin{center}
\begin{tabular}{lcc}
& Grand-canonical  ensemble & Canonical ensemble \\
$1\leq z<2$& \includegraphics[scale=0.475]{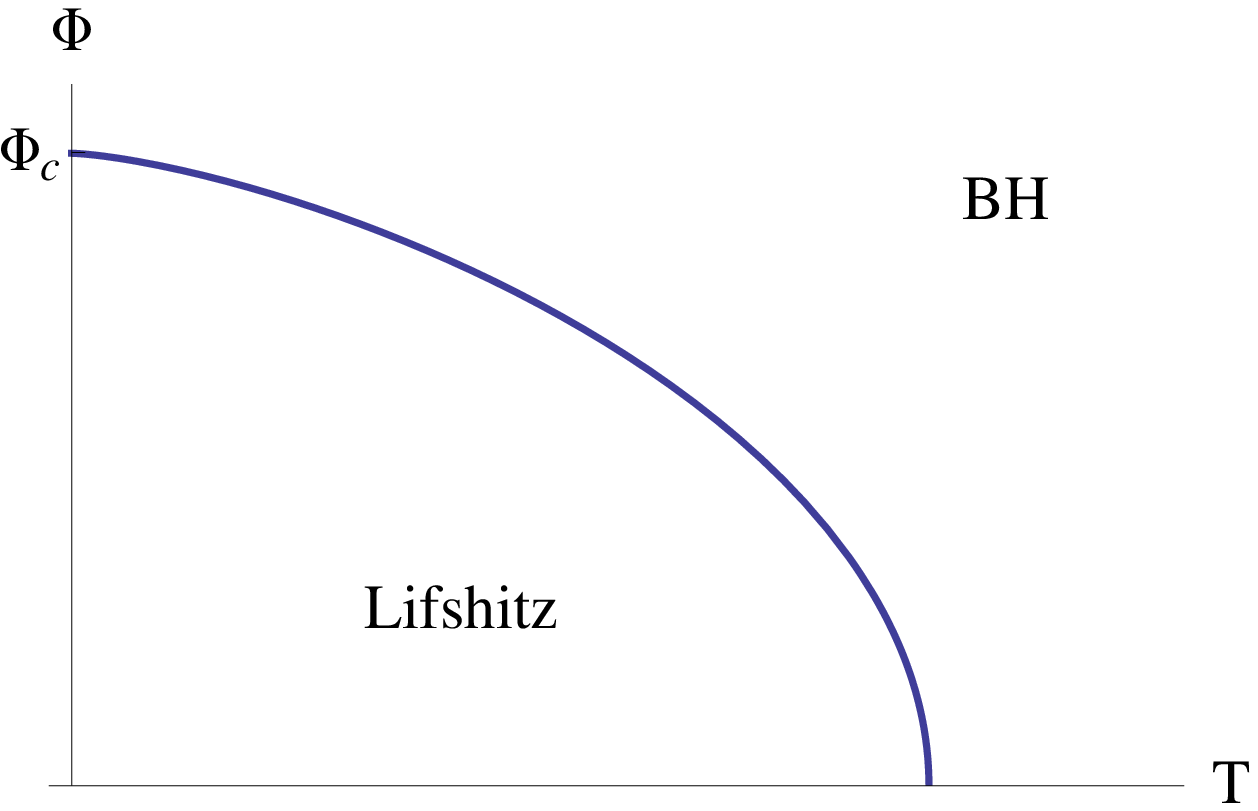} & \includegraphics[scale=0.475]{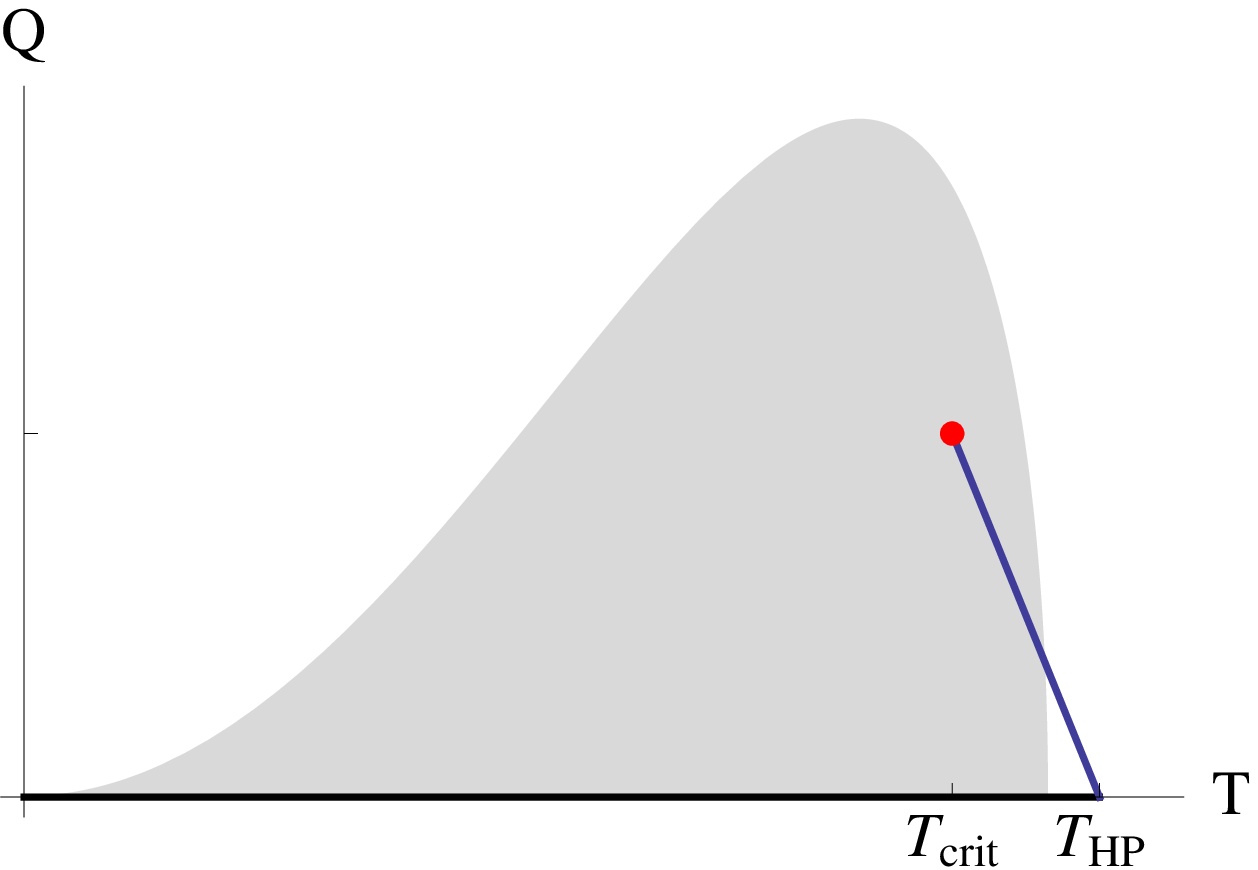} \\ 
$z=2$ & \includegraphics[scale=0.475]{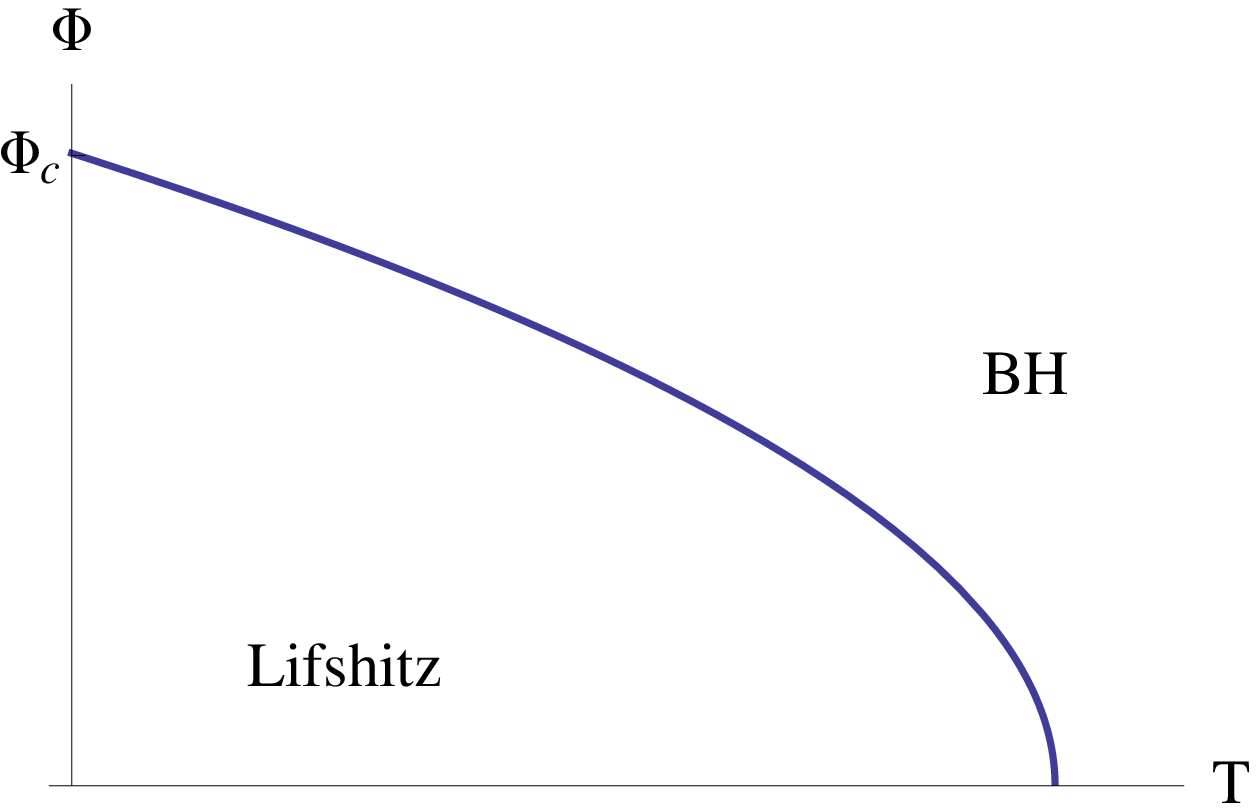} & \includegraphics[scale=0.475]{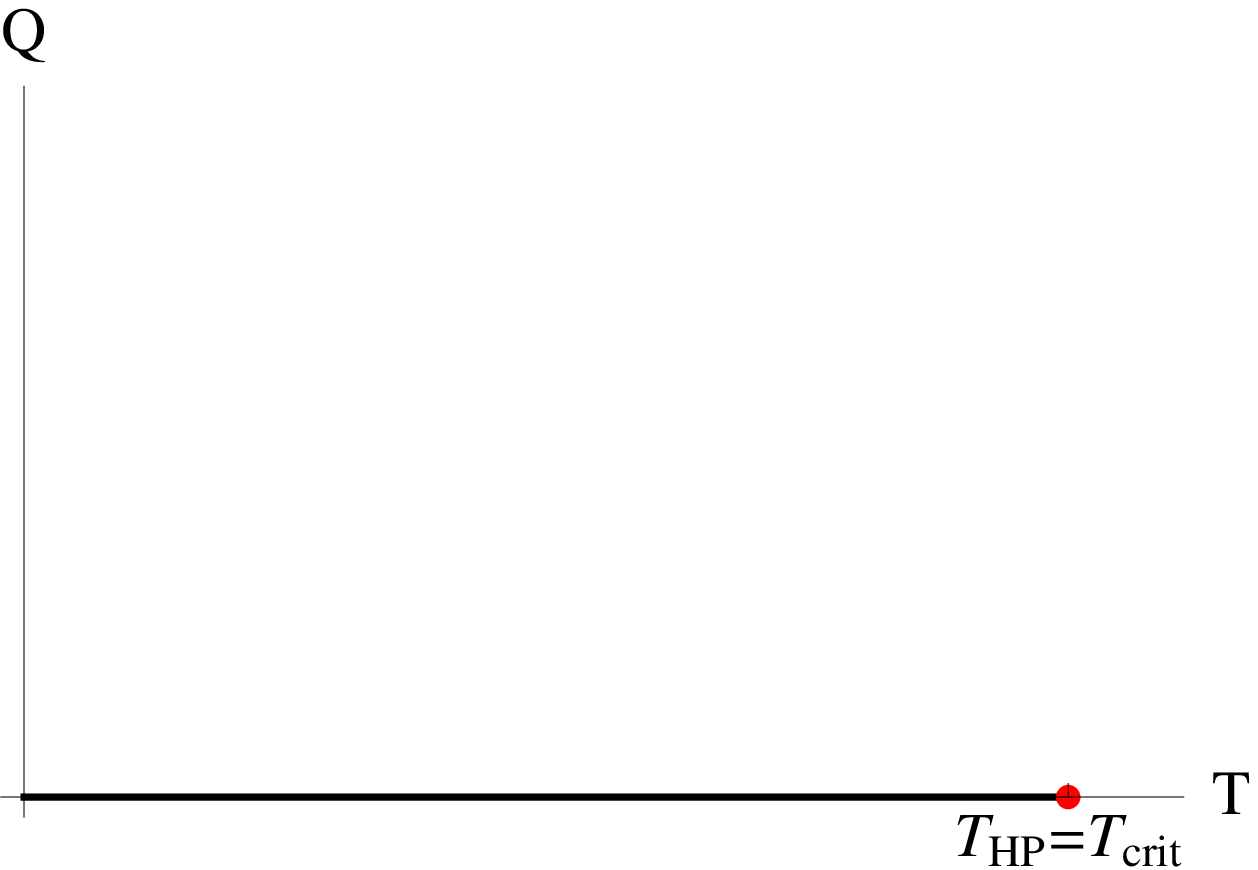} \\
$z>2$ & \includegraphics[scale=0.475]{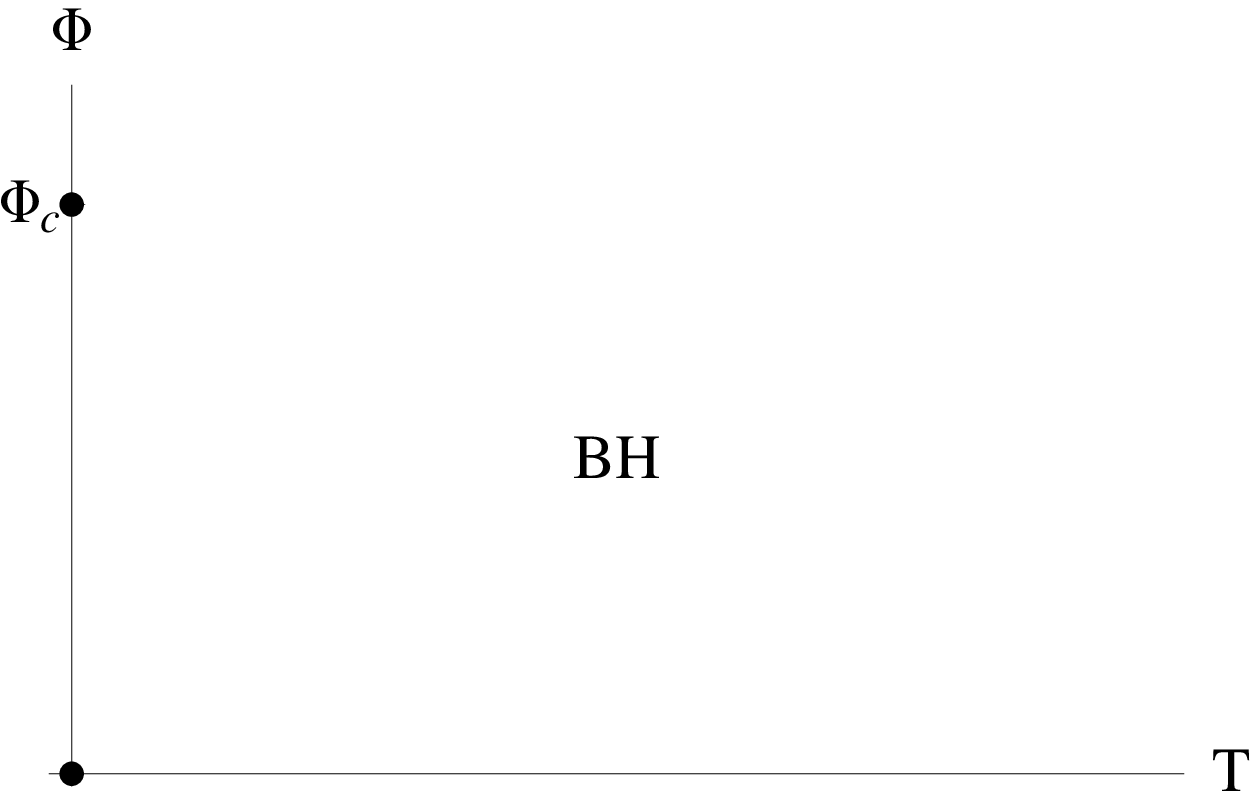} & \includegraphics[scale=0.475]{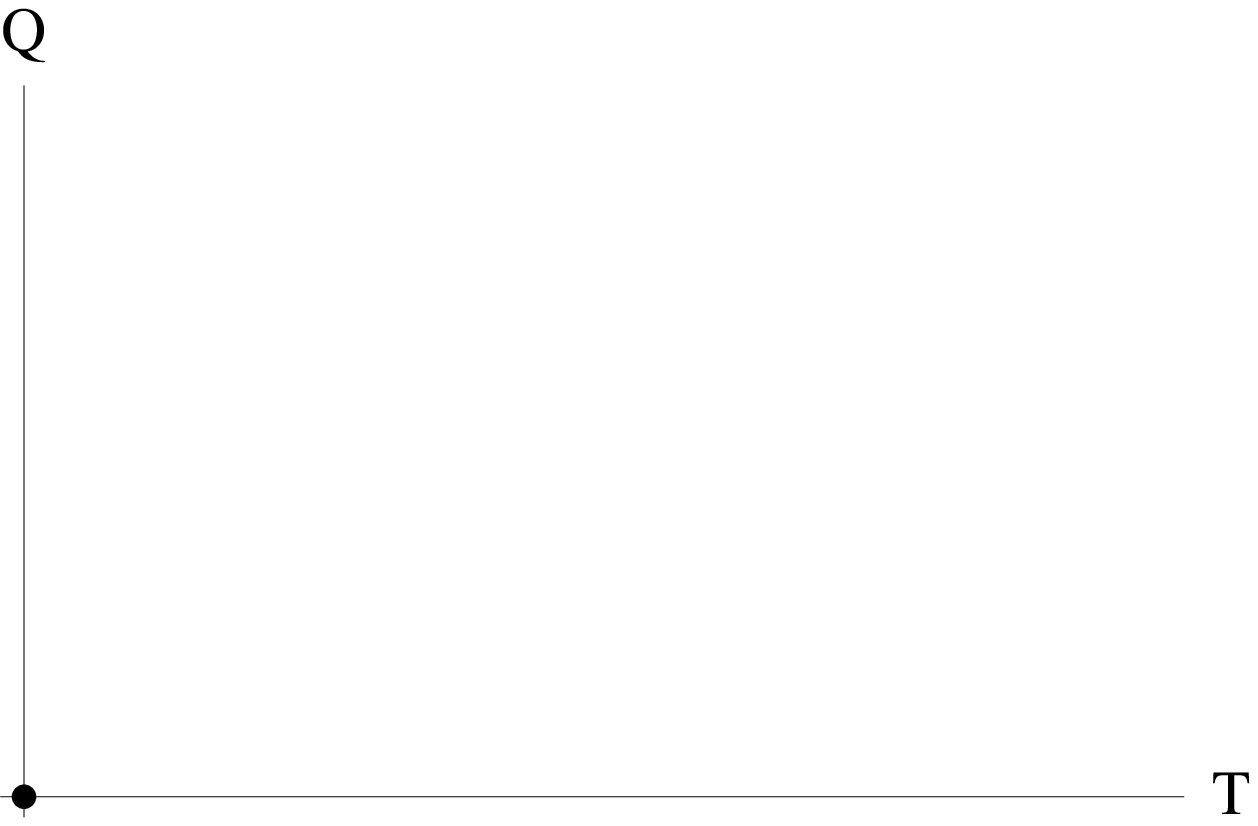} \\
\end{tabular}
\caption{Sketch of the phase diagrams obtained here for different values of the dynamic exponent $z$. See text for details.}\label{fig.phases}
\end{center}
\end{figure}

The study of the thermodynamic properties of the Lifshitz black hole solution leads to the phase diagrams showed in figure \ref{fig.phases}, both for the canonical and grand-canonical ensembles, where we sketch the results derived in the rest of the paper.

In this figure we observe that the phase transitions depend crucially on the value of the dynamical exponent $z$. For $1\leq z<2$ we find a situation completely analogous to the one studied in \cite{Chamblin:1999tk} (which corresponds to the $z=1$ AdS case).  In the grand-canonical ensemble there is a line of first order phase transitions (the blue line on the top-left diagram), where the thermodynamically preferred solution is given by thermal Lifshitz spacetime at low values of the temperature and the chemical potential and by the black hole solution in the rest of the parameter space. As in the AdS case, the $T=0$ line is dominated by Lifshitz spacetime below a critical value of the chemical potentical $ \Phi_c$ (see equation \eqref{eq.criticalpotential}), and by extremal black holes for larger values. These extremal black holes have a non-vanishing entropy, and therefore are not expected to correspond to the true ground state of the theory. In the canonical ensemble, for low values of the charge, there is a first order phase transition between small and large black holes (for low and large temperatures respectively) which ends at a critical point (blue line and red point in the top-right diagram), above which the transition between small and large black holes is smooth. However, the solutions described in this paper present an electric instability, given by the shadowed region (see \cite{Chamblin:1999hg} for the AdS case). This instability implies that the addition of an infinitesimal charge to the black hole would contribute to a reduction of the electric potential. At zero charge a Hawking-Page transition between thermal Lifshitz spacetime and a black hole occurs, whereas at $T=0$ the solution is dominated by extremal black holes.

When $z=2$ the phase diagram in the grand-canonical ensemble remains qualitatively the same as in the $1\leq z<2$ case. However, in the canonical ensemble the critical value of the charge for which there are no phase transitions above it is precisely given by $ Q=0$ (see the middle-right diagram in figure \ref{fig.phases}), and therefore only the Hawking-Page transition between thermal Lifshitz spacetimes and black holes setups remains. Furthermore, the electric instability also dissappears from the phase diagram at this value of the dynamic index.

Last, for values of the dynamical exponent $z>2$ the phase diagrams simplify even more. In the grand-canonical ensemble the line of first order phase transitions disappears, and black holes dominate the thermodynamics in almost all the parameter space. The exceptions occur at $T=0$, where the extremal black hole dominates everywhere but exactly at the point $ \Phi= \Phi_c$, where the description is given by Lifshitz spacetime. Obviously, when $ \Phi=0$ the solution is also given by Lifshitz spacetime, since it is the only description available to us. In the canonical ensemble black holes dominate the phase diagram everywhere. These situations are shown in the bottom diagrams of figure \ref{fig.phases}, where the black dots signal the points where the description is given by Lifshitz spacetime.

\subsection{Outline}

The rest of the paper is organized in the following way. We start in section \ref{sec.solution} by presenting the simple case in which the action \eqref{eq.action} has two $U(1)$ gauge fields and the solution is of the black brane type. Once this is done we proceed to consider the case of the black hole with $N\geq2$ gauge fields, and analyze the limiting cases commented on in the previous summary.

In section \ref{sec.thermo} we calculate the relevant thermodynamic quantities associated to the black holes found in section \ref{sec.solution}. These will be later used in section \ref{sec.phases} to unravel the phase diagram of the system for generic number of dimensions and dynamic index.

We conclude in section \ref{sec.remarks} with some last comments and remarks.

\section{Setup and solution}\label{sec.solution}

Consider the action \eqref{eq.action}, consisting of the usual Einstein gravity with a cosmological constant $\Lambda<0$, a scalar, and $N$ $U(1)$ gauge fields $A_i$ coupled to the scalar. The equations of motion following from it are
\beqa
 R_\mn - 
\frac{2\Lambda}{d-1}g_\mn
-\frac{1}{2} \partial_\mu \phi \partial_\nu \phi 
- \frac{1}{2} \sum_{i=1}^N e^{\lambda_i \phi}\left( (F_i){_{\mu\sigma}}(F_i){_\nu{^{\sigma}}}  - \frac{1}{2(d-1)}  F_i^2 g_\mn\right)
 &= & 0 \ ,\label{eq.einstein} \\
D_\mu \left( e^{\lambda_i\phi} F_i^\mn \right) & = &  0 \ , \label{eq.maxwell}\\
\Box \phi - \sum_{i=1}^N \frac{1}{4} \lambda_i e^{\lambda_i \phi} F_i^2 & = & 0 \ . \label{eq.scalar}
\eeqa

We are interested in finding an asymptotically Lifshitz spacetime. Furthermore, we will assume no dependence on the spatial directions, which will have the topology of ${\mathbb{R}}^{d-1}$ in the black brane case or $S^{d-1}$ in the black hole case, and we restrict to the static case in which all the fields' dependence is along the radial direction of the asymptotically Lifshitz spacetime. We now choose an ansatz for the metric based on a {\it single function} $b_k(r)$,
\be\label{eq.metricbh}
\dd s^2 =  \frac{\ell^2}{r^2} \frac{\dd r^2}{b_k(r)} - b_k(r) \frac{r^{2z}}{\ell^{2z}} \dd t^2 + r^2 \dd \Omega_{k,d-1}^2 \ ,
\ee 
with $\dd \Omega_{k,d-1}^2 $ the metric of a unit-radius $S^{d-1}$ if $k=1$ or the metric of $\mathbb{R}^{d-1}$ if $k=0$. $\ell$ is the radius of the spacetime. One can also consider the case $k=-1$, based on a hyperbolic metric for 
$\dd \Omega_{k=-1,d-1}^2 $, but later on in this paper we derive that $k=-1$ is not acceptable unless $z=1$. The function $b_k$ should asymptote $1$ at large values of the radius to recover Lifshitz spacetime. The spatial components of the gauge fields are chosen to vanish to preserve rotational symmetry, and we will work in the gauge in which the radial component is also vanishing, therefore we will be concerned only with  $A_{i,t}(r)$ turned on\footnote{To avoid confusion in our notation, let us emphasize that $A_{i,t}(r)$ denotes the temporal component of the $i-$th gauge field $A_{i,\mu}$, and not its time-derivative.}
. Finally, for the scalar we consider it to be a function  of the radial coordinate only.

\subsection{Black brane with two $U(1)$ fields}

We consider first the warmup exercise in which we set $k=0$ (constant-$r$ slices with $\mathbb{R}_t \times\mathbb{R}^{d-1}$ topology) and $N=2$ in the ansatz given before
\be
\dd s^2 =  \frac{\ell^2}{r^2} \frac{\dd r^2}{b_0(r)} - b_0(r) \frac{r^{2z}}{\ell^{2z}} \dd t^2 + \frac{r^2}{\ell^2} \dd \vec x_{d-1}^2 \ .
\ee 
 The Maxwell equations and the combination of Einstein equations $E^t_t-E^r_r$ are solved by
\beqa \label{eq.phibb}
e^\phi & = &\mu\, r^{\sqrt{2(d-1)(z-1)}}\ , \\
(F_i)_{r t} & = &  \hq_i\, r^{-(d-z)} e^{-\lambda_i \, \phi} \ , \label{eq.Fibb}
\eeqa
with $i=1,2$ and $\hq_i$ integration constants. These are related to the constants of motion associated to the gauge fields $A_i$, which enter in the action via the radial derivative, and therefore there are two conserved quantities
\be\label{eq.conschargesbb}
\frac{\delta S}{\delta A_{i,t}'} = \frac{\hq_i \ell^{z-1}}{16 \pi G_{d+1}} \ ,
\ee
which correspond to charge densities, as can be seen by calculating the total charge
\be
Q_i=\frac{1}{16\pi G_{d+1}} \int e^{\lambda_i\phi} \,{}^*F_i= \frac{V_{d-1} \hq_i\ell^{z-1}}{16\pi G_{d+1}}\ ,
\ee
with $V_{d-1}=\ell^{1-d} \int \dd^{d-1} x$ a dimensionless volume factor.

The Einstein equation in any of the spatial directions, with the expressions \eqref{eq.phibb}-\eqref{eq.Fibb} plugged in, gives a first-order differential equation for $b_0$  with solution
\be\label{eq.bksolbb}
b_0= - \frac{2 \Lambda \ell^2}{(d-1)(d+z-1)} - m\, r^{-(d+z-1)} +\frac{\ell^{2z}}{2(d-1)} \sum_{i=1}^2 \frac{\hq_i^2 \mu^{-\lambda_i }\,r^{2(1-d)-\sqrt{2(d-1)(z-1)}\lambda_i}}{d-z-1+\sqrt{2(d-1)(z-1)}\lambda_i} \ .
\ee
The integration constant $m$ will be related to the mass, as we discuss in the next section. With this solution at hand the rest of the Einstein equations, and the equation of motion for the scalar, become the algebraic equation
\be\label{eq.scalaralgebraicbb}
4\Lambda\sqrt{2(d-1)(z-1)} =  \sum_{i=1}^2  \hq_i^2 r^{-2(d-1)-\lambda_i \sqrt{2(d-1)(z-1)}}\mu^{-\lambda_i } \left[ (d-1) \lambda_i - \sqrt{2(d-1)(z-1)}  \right] \ell^{2(z-1)}\ .
\ee
The r.h.s. of \eqref{eq.scalaralgebraicbb} can be equal to the l.h.s. if we choose for the first gauge field
\be\label{eq.lambda1bb}
\lambda_1=-\sqrt{2\frac{d-1}{z-1}} \qquad \hq_1^2=-4\Lambda \mu^{\lambda_1}\ell^{2(1-z)} \frac{z-1}{d+z-2} \ .
\ee
With these values, the contribution coming from the second gauge field has to vanish, which is the case if
\be\label{eq.lambda2bb}
\lambda_2 = \sqrt{2\frac{z-1}{d-1}} \ .
\ee

Let us comment what just happened. In order to satisfy the equations of motion we had to fix not only the coupling constant $\lambda_1$, but also the charge of the first gauge field in terms of the scalar field amplitude $\mu$. As discussed in the introduction, this gauge field is needed to support the structure of (asymptotically) Lifshitz spacetime. On the other hand, the second gauge field has a free charge, which will contribute to the thermodynamic analysis as a single chemical species. It is also the term responsible of having a $b_0$ function resembling that of RN black holes. With expressions \eqref{eq.lambda1bb}-\eqref{eq.lambda2bb}, the solution reads
\beqa
\dd s^2 & = & \frac{\ell^2}{r^2}\frac{\dd r^2}{b_0} - \frac{r^{2 z}}{\ell^{2z}}b_0\, \dd t^2 + \frac{r^2}{\ell^2} \dd \vec x_{d-1}^2 \ , \label{eq.solmetricbb}\\
b_0 & = &1-m\, r^{-(d+z-1)}+ \frac{\hq_2^2\mu^{-\sqrt{2\frac{z-1}{d-1}}}\ell^{2 z}}{2(d-1)(d+z-3)} \,r^{-2(d+z-2)} \ , \label{eq.solbkbb}\\
A_{1,t}' & = & \ell^{-z}  \sqrt{2 (d+z-1)(z -1)} \, \mu^{\sqrt{\frac{d-1}{2(z-1)}}}\, r ^{d +z -2} \ , \label{eq.solAt1bb}\\
 A_{2,t}' & = & \hq_2\, \mu^{-\sqrt{2\frac{z-1}{d-1}}} \,r ^{2-d -z} \ , \label{eq.solAt2bb}\\
e^\phi & = & \mu \,r^{\sqrt{2(d-1)(z-1)}} \ , \label{eq.solscalarbb}
\eeqa
where we have used 
\be
\Lambda = -\frac{(d+z-1)(d+z-2)}{2\ell^2}\ ,
\ee
to get the right asymptotics at infinity.

It is now straightforward to check that in the uncharged limit, $\hq_2\to0$, one recovers the result in \cite{Taylor:2008tg}, whereas in the AdS limit, $z\to1$, the $A_{1,t}$ field vanishes\footnote{Actually, it vanishes provided that $e^\phi=\mu<1$. However, in the AdS limit $\lambda_1\to-\infty$ and the first $U(1)$ field decouples from the system.}. Without it, this solution is nothing but the AdS-RN solution considered in \cite{Chamblin:1999tk}. These limiting cases are the ones outlined in figure \ref{fig.bbdiagram}.

The solution presents a singularity at the origin $r=0$, where curvature invariants diverge (except when $z\to1$ and $\hq_2=0$, where the spacetime is AdS). However, the existence of an event horizon at a position $r_h\geq0$ cloaks it. The parameter $m$, which is related to the mass of the black brane, has to be positive definite, otherwise there will be no horizon and the singularity becomes a naked one. Following an argument on \cite{Hoyos:2010at}, we will impose the null energy condition $T_{\mu\nu}\xi^\mu\xi^\nu\geq0$ with $\xi^\mu=(\sqrt{g^{rr}},\sqrt{-g^{tt}},\vec0)$ a null vector. From our solution it follows that $T_{\mu\nu}\xi^\mu\xi^\nu \propto \ell^2 (R^r_r-R^t_t) = (d-1)(z-1)b_0$, so the null energy condition translates into $z\geq1$. This range of values of the dynamic index also ensures that we deal with real fields.

\subsection{Generic case}

A question that arises after the analysis performed in the previous section is whether we could have obtained a black hole solution, \emph{i.e.}, a solution with $k=1$. In this case equation \eqref{eq.scalaralgebraicbb} would have changed to \eqref{eq.scalaralgebraic} below (with $N=2$ plugged in). A solution does exist by fixing the charge of the second gauge field in a similar way as done before. The resulting black hole is similar to a Schwarzschild black hole in asymptotically Lifshitz spacetimes. 

In this section we will consider the black hole case  with $N\geq2$ gauge fields, which contains the case we just referred to. The metric is given by the ansatz \eqref{eq.metricbh} with $k=1$. As in the previous case, the Maxwell equations and a combination of the Einstein equations have as a solution for the scalar and the gauge field the expressions \eqref{eq.phibb} and \eqref{eq.Fibb}. The first difference appears in the equation of motion for $b_1$, which now reads (we keep an explicit factor $k$ for later convenience)
\be\label{eq.bksol}
b_k= \frac{k(d-2)}{d+z-3} \frac{\ell^2}{r^{2}} - \frac{2 \Lambda \ell^2}{(d-1)(d+z-1)} - m\, r^{-(d+z-1)} +\frac{\ell^{2z}}{2(d-1)} \sum_{i=1}^N \frac{\hq_i^2 e^{-\lambda_i \phi_0}\,r^{2(1-d)-\sqrt{2(d-1)(z-1)}\lambda_i}}{d-z-1+\sqrt{2(d-1)(z-1)}\lambda_i} \ .
\ee
With this, the algebraic equation equivalent to \eqref{eq.scalaralgebraicbb} is
\beqa\label{eq.scalaralgebraic}
0 & = &  \sum_{i=1}^N  \hq_i^2 r^{-2(d-1)-\lambda_i \sqrt{2(d-1)(z-1)}}e^{-\lambda_i \phi_0} \left[ (d-1) \lambda_i - \sqrt{2(d-1)(z-1)}  \right] \ell^{2(z-1)}\nonumber\\
&& \quad + 2\sqrt{2(d-1)(z-1)} \left( k (d-1)(d-2)r^{-2} - 2 \Lambda \right)\ .
\eeqa
As before, the first gauge field can be used to cancel the term proportional to $\Lambda$ by choosing \eqref{eq.lambda1bb}, and the contribution from the next $N-2$ gauge fields can be cancelled by choosing
\be
\lambda_j = \sqrt{2\frac{z-1}{d-1}} \ , \qquad j=2,\cdots,N-1\ .
\ee
Finally, the $N$-th gauge field can be used to cancel the term proportional to $k$ if one fixes
\be
\lambda_N=-\frac{d-2}{d-1}\sqrt{2\frac{d-1}{z-1}} \qquad \hq_N^2=k\, \mu^{\lambda_2} \ell^{2(1-z)} \frac{2(d-1)(d-2)(z-1)}{d+z-3}\ .
\ee
Notice that, once again, the charge of the first gauge field is fixed to support the Lifshitz spacetime. Additionaly, the $N$-th gauge field's charge is also fixed, in this case to support  the existence of the $S^{d-1}$ topology for $k=1$. The hyperbolic case, $k=-1$, leads to imaginary charge densities (unless $z=1$). Therefore, here and below, we will only consider the cases $k=0$ or $k=1$.

Summarizing, the generic solution we found to \eqref{eq.action} is
\beqa
\dd s^2 & = & \frac{\ell^2}{b_k}\frac{\dd r^2}{r^2} - \frac{r^{2 z}}{\ell^{2z}}b_k\, \dd t^2 + r^2 \dd \Omega_{k,d-1}^2 \ , \label{eq.solmetric}\\
b_k & = &k \left( \frac{d-2}{d+z -3} \right)^2 \frac{\ell^2}{r^2} +1-m\, r^{-(d+z-1)}+\sum_{j=2}^{N-1}\frac{\hq_j^2\, \mu^{-\sqrt{2\frac{z-1}{d-1}}}\ell^{2 z}}{2(d-1)(d+z-3)} \,r^{-2(d+z-2)} \ , \label{eq.solbk}\\
A_{1,t}' & = & \ell^{-z}  \sqrt{2 (d+z-1)(z -1)} \, \mu^{\sqrt{\frac{d-1}{2(z-1)}}}\, r ^{d +z -2} \ , \label{eq.solAt1}\\
A_{j,t}' & = & \hq_j\, \mu^{-\sqrt{2\frac{z-1}{d-1}}} \,r ^{2-d -z} \ , \qquad \qquad \qquad ( j=2,\cdots,N-1) \,\label{eq.solAt2}\\
A _{N,t}' & = & \ell^{1-z}  \frac{\sqrt{2k(d -1) (d -2) (z -1)} }{\sqrt{d +z -3}}  \mu^\frac{(d-2)}{\sqrt{2(d-1)(z-1)}}\, r ^{d +z -4}\ , \label{eq.solAt3}\\
e^\phi & = & \mu \,r^{\sqrt{2(d-1)(z-1)}} \ , \label{eq.solscalar}
\eeqa
where again $\Lambda = -(d+z-1)(d+z-2)/2\ell^2$.

We see that this solution depends on the parameters $\hq_j$, $\mu$ and $m$, which will correspond to the charge densities, the amplitude of the scalar field and the energy of the black hole, respectively. As commented several times already, two of the gauge fields have their charges fixed to support a spherical black hole in Lifshitz spacetime. The metric presents a horizon which, in general, has a near-horizon geometry given by the direct product of a $2$--dimensional Rindler spacetime  (the coordinates being $r$ and $t$) and the spacetime given by $\dd \Omega_{k,d-1}$, which are spectator coordinates in this approximation. For a certain value of $m$ and $\rho$ the black hole becomes extremal. In this case the near-horizon geometry is given by $AdS_2\times S^{d-1}$ or $AdS_2\times {\mathbb{R}}^{d-1}$ depending on whether $k=1,0$. 

We can take the AdS limit $z\to1$. In this case we recover once again the results in \cite{Chamblin:1999tk}. The fixed-charge gauge fields $A_{1,t}$ and $A_{N,t}$ are set to zero (provided $\mu<1$) and their coupling to the scalar field in the action goes to $\lambda_{1,N}\to-\infty$, decoupling them from the rest of the matter fields. The scalar field becomes constant as well. Figure \ref{fig.bhdiagram} represents this limit and the reduction of the number of charges in consideration.

We are also interested in recovering the solution characterized by $k=0$ (this is, with flat topology of the constant-$r$ slices) as an explicit limit of \eqref{eq.solmetric}-\eqref{eq.solscalar}. Following \cite{Chamblin:1999tk}, we introduce the dimensionless parameter $\eta$ and we scale $r\to \eta\, r$. Given the form of the metric  \eqref{eq.solmetric} we must impose also $t\to \eta^{-z} t$. We will focus on the neighborhood of a point in the $S^{d-1}$, considering just a flat metric around it $\ell^2 \dd \Omega_{1,d-1}^2 \to \eta^{-2}\dd \Omega_{0,d-1}^2$. Taking the scalar field to be scale-invariant, the appropriate scaling is fixed to be $\mu\to\eta^{-\sqrt{2(d-1)(z-1)}} \mu$. Now, the scaling on $\mu$ determines completely the scaling of the $A_{1,t}$ and $A_{N,t}$ gauge field charges. For the remaining $\hq_j$ charges we use the fact that these parameters actually correspond to charge densities, and therefore, to have scale-invariant charge, they have to transform under the scaling as $\hq_j \to \eta^{d-1} \hq_j$. All in all, the $N$ field strengths scale as
\be
F_1   \to F_1  \ , \quad F_j  \to F_j \ , \quad F_N   \to \eta^{-1}F_N \,
\ee
so in the large $\eta\to\infty$ limit we should not consider  the gauge field $A_N$, consistently with setting $k=0$ in the general solution. Furthermore, in this limit the metric becomes \eqref{eq.solmetricbb} with a blackening function $b_0$ given by \eqref{eq.solbkbb} (there $N=2$). So we consistently reduced the black hole case to the black brane one, and are able to construct the web of relations depicted in figure \ref{fig.bbbhdiagram}.

An interesting special class of solutions (for $k=1$ and $N=3$ for simplicity) are those that satisfy the mass-charge relation
\be\label{BPS-mass}
m^2 = \frac{2(d-2)^2\ell^{2(1+z)}\mu^{-\sqrt{2\frac{z-1}{d-1}}}}{(d-1)(d+z-3)^3}\rho^2 \ .
\ee
In this case, the black function reduces to
\be\label{BPS-BH}
b_1=1+\left( \frac{d-2}{d+z-3}\frac{\ell}{r} - \frac{\rho \mu^{-\frac{1}{2}\sqrt{2\frac{z-1}{d-1}}}\ell^z}{\sqrt{2(d-1)(d+z-3)}} r^{-(d+z-2)}\right)^2\ .
\ee
This function is always positive and therefore there is no horizon. At $r=0$ there is instead a naked singularity. 
For $z=1$ this is precisely what happens for the supersymmetric RN solutions. For $z\neq 1$, one may expect that the solution with \eqref{BPS-mass} and \eqref{BPS-BH} can be embedded as BPS solutions in some gauged supergravity action. 
These would need to be extensions of the present action \eqref{eq.action}, since in gauged supergravity with scalar fields, the scalar potential is not constant.

Lastly, notice that even when the case $d=2$ is not included in our analysis, it turns out that the solution also works in this case. For this value of the dimension, the solutions with $k=0$ and $k=1$ coincide locally. This is so because in this case there is only one spatial direction, and the difference between the two solutions is a global matter: whether the direction is compact or not.

\section{First law of thermodynamics} \label{sec.thermo}

We will discuss now the thermodynamic properties associated to the solution presented in \eqref{eq.solmetric}-\eqref{eq.solscalar}. For the sake of clarity we will restrict to $3$ $U(1)$ gauge fields with a black hole with spherical topology, \emph{i.e.}, just one non-trivial charge $\hq_2\equiv\hq$. We will mark the difference between the black hole and black brane analysis keeping explicit terms of $k$, though.

\paragraph{Temperature and entropy}

Unfornutately, it is not possible to obtain  a general, analytical expression for the position of the horizon $r_h$ (given by the larger positive root of $b_k(r_h)=0$) as a function of the three parameters $m$, $\mu$ and $\hq$. However, we can still proceed to a thermodynamic study. 

Let us first notice that even when the parameter $m$ is related to the mass of the black hole, as we will show in \eqref{eq.mass}, it is not a fundamental parameter of the theory. These parameters are given by $\mu$,  the amplitude of the scalar, $\hq$, corresponding to the charge (potential) in the canonical (grand-canonical) ensemble and the temperature $T$. Therefore we find it convenient to express $m=m(\mu,\hq,T)$. However, as it is not possible either to obtain a closed expression for the temperature, we will use instead $m=m(\mu,\hq,r_h)$, given by
\be\label{eq.hmexpression}
m=r_h^{d+z-1} \left[ 1+k\left(\frac{d-2}{d+z-3}\right)^2\frac{\ell^2}{r_h^2} + \frac{\hq^2 \mu^{-\sqrt{2\frac{z-1}{d-1}}}\ell^{2z}}{2(d-1)(d+z-3)}r_h^{-2(d+z-2)} \right]\ .
\ee
Notice that $m$ is non-negative. Using this expression,  the temperature as a function of the position of the horizon radius $r_h$, $\mu$ and $\hq$ reads
\be\label{eq.temperature}
T=\frac{r_h^z}{4\pi \ell^{1+z}} \left[ (d+z-1) + k \frac{(d-2)^2}{d+z-3} \frac{\ell^2}{r_h^2} - \frac{\hq^2\mu^{-\sqrt{2\frac{z-1}{d-1}}}\ell^{2z}}{2(d-1)} r_h^{-2(d+z-2)} \right] \  .
\ee
The entropy is given, as usual, by the Bekenstein-Hawking formula
\be
S = \frac{V_{d-1}}{4 G_{d+1}} r_h^{d-1} \ .
\ee

The temperature \eqref{eq.temperature} vanishes when the horizon radius satisfies
\be
\hq^2 = 2(d-1)\mu^{\sqrt{2\frac{z-1}{d-1}}}  \ell^{-2z} \left( (d+z-1) + k \frac{(d-2)^2}{d+z-3} \frac{\ell^2}{r_{ext}^2} \right) r_{ext}^{2(d+z-2)}  \ ,
\ee
where we have denoted with $r_{ext}$ the position of the horizon at extremality, defined by the conditions
$b(r_{ext})=b'(r_{ext})=0$. Using the relation \eqref{eq.hmexpression} we can determine the value of $m_{ext}$ as
\be\label{eq.mext}
m_{ext} =\left[k \frac{2(d-2)^2 \ell^2}{(d+z-3)^2}\frac{ r_{ext}^{d+z-3}}{d+z-1}+ \frac{d+z-2}{d+z-3} \frac{\hq^2 \mu^{-\sqrt{2\frac{z-1}{d-1}}} \ell^{2z}\, r_{ext}^{3-d-z}}{(d-1)(d+z-1)} \right]  \ .
\ee

\paragraph{Mass}

To evaluate the mass we should study the renormalized one point functions, where a counterterm must be added to regularize the expressions at infinity, since divergences will appear. However, here we wil take an alternative approach. We decide to evaluate the Komar integral
\be
M_T = - \frac{1}{8\pi G_{d+1}} \oint \dd S_{\mn}D^\mu K_T^\nu \ ,
\ee
with $K_T=\partial_t$ and substract the result from the thermal case (this is, the case with $m=\hq=0$, but such that the euclideanized time is periodic). We must match the normalizations of the Killing vectors between the black hole and thermal cases at $r_\infty$ to ensure they have the same norm expression at infinity. This is done by considering
\be \label{eq.betamatch}
K_0 = \frac{\sqrt{b(r_\infty)}}{\sqrt{b_0(r_\infty)}} K_T \ ,
\ee
with $K_0$ the Killing vector in the thermal setup. The result is
\be\label{eq.mass}
M= M_T-M_0  = \frac{V_{d-1}}{16\pi G_{d+1}} m \, \ell^{-1-z} (d-1) \ ,
\ee
where $V_{d-1}$ is the volume of the unit $S^{d-1}$ sphere.

Expression \eqref{eq.mass} will be useful in the grand-canonical case, where the charge is free to vary but the potential is fixed. In the canonical ensemble, however, one must fix the charge, which is proportional to $\hq$, and therefore the correct comparison scheme is to substract the result of the extremal black hole with the appropriate value for $r_{ext}$. It is not difficult to show that in this case
\be
\Delta M= M_T-M_{ext}  = \frac{V_{d-1}}{16\pi G_{d+1}} (m-m_{ext})\ell^{-1-z} (d-1) \ .
\ee

\paragraph{Charges and chemical potential}

As commented above, there are three conserved charges
\be
Q_i= \frac{V_{d-1} \hq_i \ell^{z-1}}{16\pi G_{d+1}}\ ,
\ee
but two of them are completely specified in terms of the metric and the scalar
\beqa
Q_1 &= &  \frac{V_{d-1} \ell^{-1}}{16\pi G_{d+1}} \sqrt{2(d+z-1)(z-1)} \mu^{-\sqrt{\frac{d-1}{2(z-1)}}} \ , \\ 
   Q_2 & = &  \frac{V_{d-1} \ell^{z-1}\hq}{16\pi G_{d+1}}\ , \\
  Q_3 & = & k \frac{V_{d-1}}{16\pi G_{d+1}} \sqrt{\frac{2(d-1)(d-2)(z-1)}{d+z-3}}   \mu^{-\frac{d-2}{d-1}\sqrt{\frac{d-1}{2(z-1)}}}   \ .
\eeqa

The potentials associated to these charges in the thermodynamic relations come from the form of the  the gauge fields potentials as functions of the radial coordinate
\beqa
A_{1,t} & = &\sqrt{\frac{2(z-1)}{d+z-1}} \mu^{\sqrt{\frac{d-1}{2(z-1)}}}  \ell^{-z} \left(r^{d+z-1}-r_h^{d+z-1}\right) \ , \\
 A_{2,t} & = & - \frac{\hq\, \mu^{-\sqrt{2\frac{z-1}{d-1}}} }{d+z-3} \left( r^{3-d-z}-r_h^{3-d-z}\right) \ , \\
A_{3,t} & = & k \frac{\sqrt{2(d-1)(d-2)(z-1)}}{(d+z-3)^{3/2}} \mu^{\frac{d-2}{\sqrt{2(d-1)(z-1)}}} \ell^{1-z} \left( r^{d+z-3}-r_h^{d+z-3} \right) \ , 
\eeqa
where we have fixed the integration constants such that the gauge fields vanish at the horizon, impliying that their norm-squared is non-singular there. The $A_1$ and $A_3$ fields diverge at the boundary, however they will not be of importance for a thermodynamic analysis, as we will see shortly.

\paragraph{First law of thermodynamics}

It is now straightforward to check that the first law of thermodynamics holds\footnote{From now on we will denote $Q_2\equiv Q$, since this is the only charge of importance in the thermodynamic relations.}
\be
\dd M = T \dd S +  \Phi \dd  Q \ , 
\ee
where we define 
\be
\Phi =   A_{2,t}(\infty) = \frac{\hq \mu^{-\sqrt{2\frac{z-1}{d-1}}}}{d+z-3} r_h^{3-d-z} \ .
\ee
 Specifically we have
\be
T = \left( \frac{\partial M}{\partial S} \right)_{ Q} \ , \quad  \Phi = \left( \frac{\partial M}{\partial  Q} \right)_{S} \ .
\ee

In case we compare to the extremal case and not the thermal one, we have to consider the thermodynamic relation
\be
\dd (\Delta M )=  T \dd S+ \left(  \Phi- \Phi_{ext} \right) \dd  Q \ .
\ee

We can calculate now the heat capacity at constant charge, finding the result
\beqa\label{eq.heatcap}
C_{ Q} & = & T \left( \frac{\partial S}{\partial T} \right)_{ Q} = \frac{\partial M / \partial r_h}{\partial T/ \partial  r_h}\\
& = & \frac{\pi\, T\,V_{d-1}}{G_{d+1}} \frac{(d-1) \ell^{1+z} r_h^{d-z-1}}{z(d+z-1) + k  \frac{(d-2)^2(z-2)}{d+z-3} \frac{\ell^2}{r_h^2} + \frac{2d+z-4}{2(d-1)} \hq^2 \mu^{-\sqrt{2\frac{z-1}{d-1}}} \ell^{2z}r_h^{2(2-d-z)}} \nonumber \ ,
\eeqa
from where we conclude that the specific heat at constant charge is always positive and regular for $z\geq 2$. For $1\leq z <2$ there is an instability for some values of the black hole parameters in the spherically symmetric case. We will comment further on this in the next section.

Notice also that, for the first law of thermodynamics to be satisfied, the gauge fields $A_1$ and $A_3$ (and their associated charges) are not needed. The fact that these two gauge fields do not seem to affect the thermodynamics may be related to having their charges completely determined by the scalar parameter $\mu$ and their diverging at the boundary, which would affect drastically a potential holographic interpretation. Therefore, it seems natural to assume that these fields, which are needed just to support the structure of the asymptotically Lifshitz spacetime solution, do not have a thermodynamic interpretation.

\section{Phase structure} \label{sec.phases}

\subsection{Grand-canonical ensemble}

Let us  define the free energy from the thermodynamic relation
\be
W= M - TS -  \Phi  Q \ ,
\ee
where we have not included the contribution coming from $\Phi_1 Q_1+\Phi_3 Q_3$, \emph{i.e.}, as these two charges are fixed to support the asymptotic topology of the black hole solution, they must correspond to an ensemble in which those terms do not contribute to the free energy. In other words, we keep the charges $Q_1$ and $Q_3$ fixed, since otherwise these gauge fields would spoil the asymptotic topology of our solution. This is equivalent to keep the value of $\mu$ fixed in the phase structure analysis performed in the following.

Defining a critical potential given by
\be\label{eq.criticalpotential}
 \Phi_c^2 = k \frac{2(d-1)(d-2)^2}{(d+z-3)^3} \ell^{2(1-z)} \mu^{-\sqrt{2\frac{z-1}{d-1}}} \ ,
\ee
we can write parametric equations for the temperature $T$ and the free energy $W$
\beqa
\label{eq.grandcanonicalF}
W & = & \frac{V_{d-1} \ell^{-1-z}}{16\pi G_{d+1}} r_h^{d+z-1}  \left[  -z + (2-z) \frac{d+z-3}{2(d-1)} \mu^{\sqrt{2\frac{z-1}{d-1}}} \left( \Phi_c^2  -  \Phi^2\right) \ell^{2z} r_h^{-2}\right] \ , \\
T & = & \frac{r_h^z \ell^{-1-z}}{4\pi} \left[ (d+z-1)  + \frac{ (d+z-3)^2}{2(d-1)} \mu^{\sqrt{2\frac{z-1}{d-1}}} \left(  \Phi_c ^2 -  \Phi^2 \right)  \ell^{2z}  r_h^{-2} \right] \  . \label{eq.grandcanonicalT}
\eeqa

\paragraph{The $z=2$ case}

Let us start the analysis of the parametric equations \eqref{eq.grandcanonicalF} and \eqref{eq.grandcanonicalT} by studying the case with $z=2$. The temperature will be a bijective function of the radius of the horizon, and therefore every temperature is described by only one black hole. Clearly, there is a minimum value for $T$ given by the setup with a vanishing radius of the horizon
\be\label{eq.Tz2}
T_{(r_h=0,z=2)} = \frac{(d-1)}{8\pi} \mu^{\sqrt{\frac{2}{d-1}}} \ell \left( \Phi_c^2 - \Phi^2  \right) \ .
\ee
The previous expression is negative when $\Phi^2 >  \Phi_c^2$, and a quick look at equation \eqref{eq.grandcanonicalT} shows that the temperature  diverges as  $r_h\to\infty$. Therefore, being the relation between $T$ and $r_h$ a bijective one, all positive values of the temperature are supported by a black hole. That this is the thermodynamically preferred solution is clear from the expresion for the free energy as a function of $T$ and $ \Phi$
\be
W = -\frac{V_{d-1}}{8\pi G_{d+1}}  \left(\frac{4\pi}{d+1}\right)^\frac{d+1}{2} \ell^{\frac{3}{2}(d-1)} \left( T- T_{(r_h=0,z=2)} \right)^\frac{d+1}{2} \ .
\ee
 On the contrary, for potentials less than the critical potential (in absolute value) the low temperature description has to be given by thermal Lifshitz spacetime. The line along which this phase transition occurs can be given in analytic form
 \be
 \Phi= \frac{\sqrt{2}\, \mu ^{-\sqrt{\frac{1}{2}\frac{1}{d-1}}}}{(d-1)\ell} \sqrt{(d-2)^2-4 \pi  (d-1) T \,\ell} \sim \left( T_c^{(z=2)} -T\right)^{1/2} \ ,
 \ee
 where $T_c^{(z=2)}=(d-2)^2/4\pi(d-1) \ell$.

\paragraph{The $1\leq z<2$ case}

In this case, the competition between the two terms in the free energy \eqref{eq.grandcanonicalF} depends on the sign of $ \Phi_c^2 -  \Phi^2$. This case is reminiscent of the $z=1$ case studied in \cite{Chamblin:1999tk}.

\begin{figure}[tb]
\begin{center}
\includegraphics[scale=0.7]{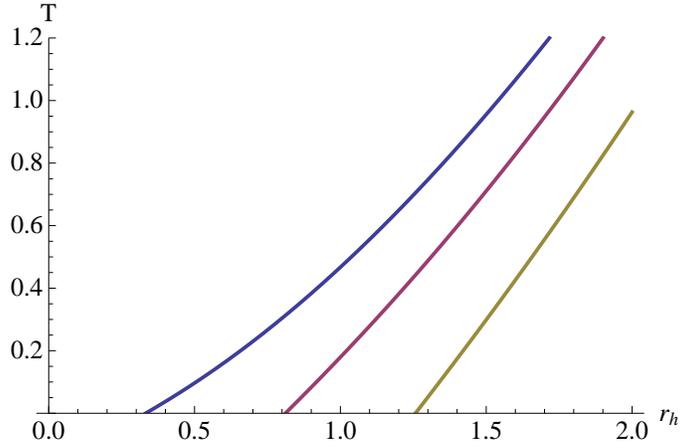}
\caption{ Temperature as a function of the horizon radius for $ \Phi=1.1  \Phi_c$ (blue), $ \Phi=1.5  \Phi_c$ (red) and $ \Phi=2  \Phi_c$ (yellow). There is a minimum  radius at which $T=0$. An evaluation of the free energy shows that it is negative everywhere along the lines, specifically at $T=0$. This plot was made for $d=6$, $z=1.6$ and $\mu=2$ in units where $\ell=1$.}\label{fig.largephi}
\end{center}
\end{figure}

For values of the potential larger than the critical one the free energy is clearly negative and the black hole setup is favored in all the temperatures where such a description is valid. Now, in view of \eqref{eq.grandcanonicalT}, it is clear that there are two terms: one positive and proportional to $r_h^z$ and one negative and proportional to $r_h^{z-2}$. There will be a positive value of the horizon radius at which the temperature vanishes, and above this value the temperature will be a monotonically increasing function of the radius (see figure \ref{fig.largephi}). These two results imply that, for values of the potential greater than the critical one, there exists a black hole description of the system which, in turn, is thermodynamically preferred. Indeed, when $T=0$ at a finite radius of the horizon and the potential is not at its critical value $\Phi_c$, the free energy is given by
\be \label{eq.extremalW}
W = - \frac{V_{d-1} \ell^{-1-z}}{8\pi G_{d+1}}  \frac{d-1}{d+z-3} r_h^{d+z-1} \ ,
\ee
showing that in this case the system at zero temperature is described by an extremal black hole, since $W<0$ when $r_h>0$.

\begin{figure}[tb]
\begin{center}
\includegraphics[scale=0.7]{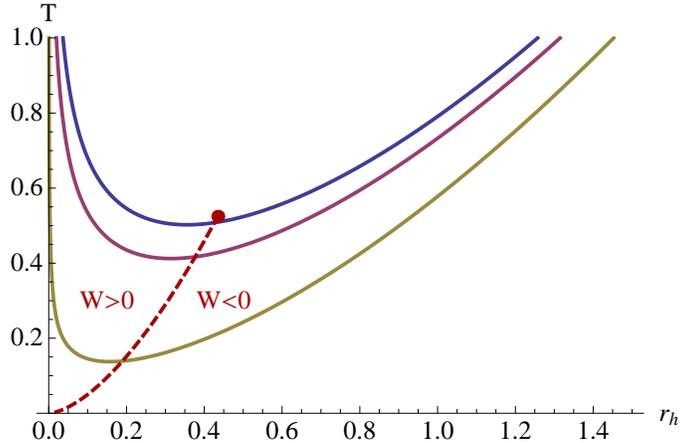}
\caption{Temperature as a function of the horizon radius for $ \Phi=0.2  \Phi_c$ (blue), $ \Phi=0.5  \Phi_c$ (red) and $ \Phi=0.9  \Phi_c$ (yellow). There is a minimum temperature the black holes can describe. The dashed line marks the position at which the evaluation of the free energy along the lines changes sign, and it finishes just above the blue line, on the graph for $\Phi=0$ (not plotted). For small black holes the free energy is positive whereas for the larger ones it is negative. The values used to produce this plot were $d=6$, $z=1.6$ and $\mu=2$ in $\ell=1$ units.}\label{fig.lowphizl2}
\end{center}
\end{figure}

When the potential $ \Phi$ is less than the critical value the expression for the temperature \eqref{eq.grandcanonicalT} is always positive and diverges when $r_h\to0$ and $r_h\to\infty$. When the horizon radius for a given potential is given by $r_h^2=\frac{(2-z)(d+z-3)^2}{2 z (d-1) (d+z-1)} \mu^{\sqrt{2\frac{z-1}{d-1}}} \ell^{2z} ( \Phi_c^2 -  \Phi^2)$, the temperature approaches a minimum value
\be\label{eq.grandcanonicalline}
T_{min}  =  \frac{d+z-1}{2\pi (2-z)} \ell^{-1-z} \left[ \frac{(2-z)(d+z-3)^2}{2z(d-1)(d+z-1)} \mu^{\sqrt{2\frac{z-1}{d-1}}} \ell^{2z} \left(  \Phi_c^2 -  \Phi^2 \right)  \right]^{z/2} \ .
\ee
 For any temperature larger than this there are two possible horizon radii. When we study the free energy, it is direct to see that $W(T_{min})>0$. Two branches depart from this point, one corresponding to values of the horizon approaching $r_h=0$ and one approaching $r_h=\infty$, \emph{i.e.}, small and large black holes respectively. The branch corresponding to the small black holes gives always a positive free energy, whereas the branch associated to the large black holes corresponds to a negative free energy for all values of the temperature $T>T_{min}$ (see figure \ref{fig.lowphizl2}). Therefore, there is a first order phase transition between a thermal Lifshitz spacetime and a black hole when $ \Phi^2< \Phi_c^2$.

\paragraph{The $z>2$ case}

For values of the dynamic index larger than $2$ and potentials less than the critical one (in absolute value), the free energy is negative for any value of the horizon radius, as can be seen directly from \eqref{eq.grandcanonicalF}. Analyzing the expression for the temperature we observe that there are two positive terms multiplied by a positive power of $r_h$. Therefore, the temperature is a bijective function of the horizon radius, with vanishing temperature for zero horizon radius. Altogether this means that for  low potentials and $z>2$ the black-hole solution dominates the phase diagram, even at zero temperature. 

On the other hand, if $ \Phi^2 >  \Phi_c^2$ the temperature will vanish at a finite horizon radius, in a  similar way to the one reported on figure \ref{fig.largephi}. Equation \eqref{eq.extremalW} is still valid, showing that at low temperatures the thermodynamically favored description is given by the solution with a black hole. Indeed, this result holds true for any value of the temperature.

\paragraph{The $ \Phi^2= \Phi_c^2$ case}

In all the previous cases we studied the phase diagrams for values of the potential above and below $ \Phi_c$. When the potential is tuned to precisely its critical value the free energy can be expressed as a function of the temperature only
\be\label{eq.grandcanT0}
W = -z \frac{V_{d-1} \ell^{-1-z}}{16\pi G_{d+1}} \left(  \frac{4\pi T \ell^{1+z}}{d+z-1}  \right)^\frac{d+z-1}{z} \ .
\ee
At $T=0$ the preferred phase is the one described by the Lifshitz spacetime with finite potential. At any other temperature the black-hole description is the favored one. This signals the point $T=0$, $ \Phi^2= \Phi_c^2$ as a special one, since its description is always given by Lifshitz spacetime, independently of the value of the index $z$.

\subsection{Canonical ensemble}

We now proceed to study the case in which we keep the charge $Q\propto \rho$ fixed, corresponding to the canonical ensemble. In this case the free energy is defined as
\beqa
F & = & \Delta M -  TS \nonumber\\
& = & \frac{V_{d-1} \ell^{-1-z}}{16\pi G_{d+1}} \Big[ - m_{ext}(d-1) -z r_h^{d+z-1} +k \frac{(d-2)^2(2-z)}{(d+z-3)^2} \ell^2 r_h^{d+z-3}\nonumber\\
&& \qquad\qquad\quad +\frac{2d+z-4}{2(d-1)(d+z-3)}\hq^2 \mu^{-\sqrt{2\frac{z-1}{d-1}}} \ell^{2z} r_h^{3-d-z} \Big]\ ,\label{eq.canfreeen}
\eeqa
with $m_{ext}$ given in expression \eqref{eq.mext}. As in the grand-canonical case, we should investigate separately the $1\leq z<2$, $z=2$ and $z>2$ cases.

\paragraph{The $1\leq z<2$ case}

\begin{figure}[tb]
\begin{center}
\subfigure{\includegraphics[scale=0.66]{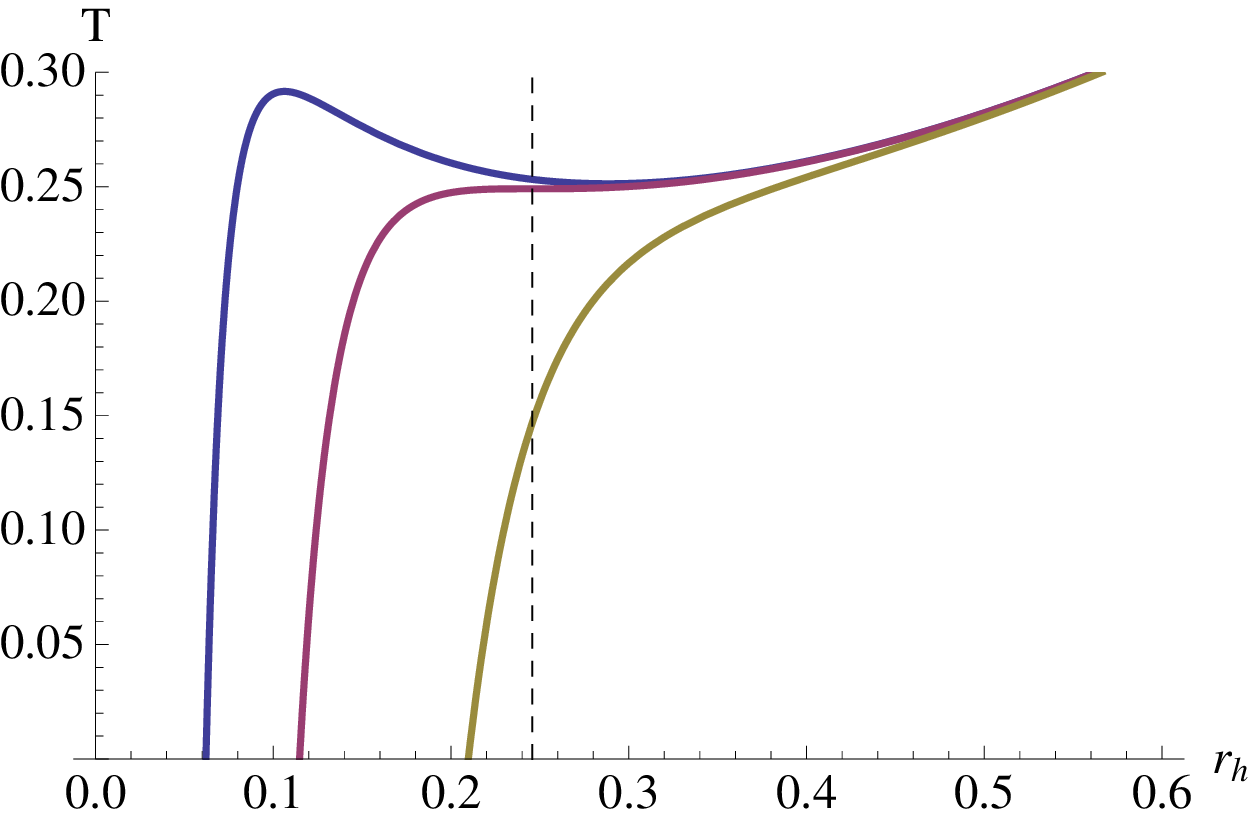}}
\subfigure{\includegraphics[scale=0.66]{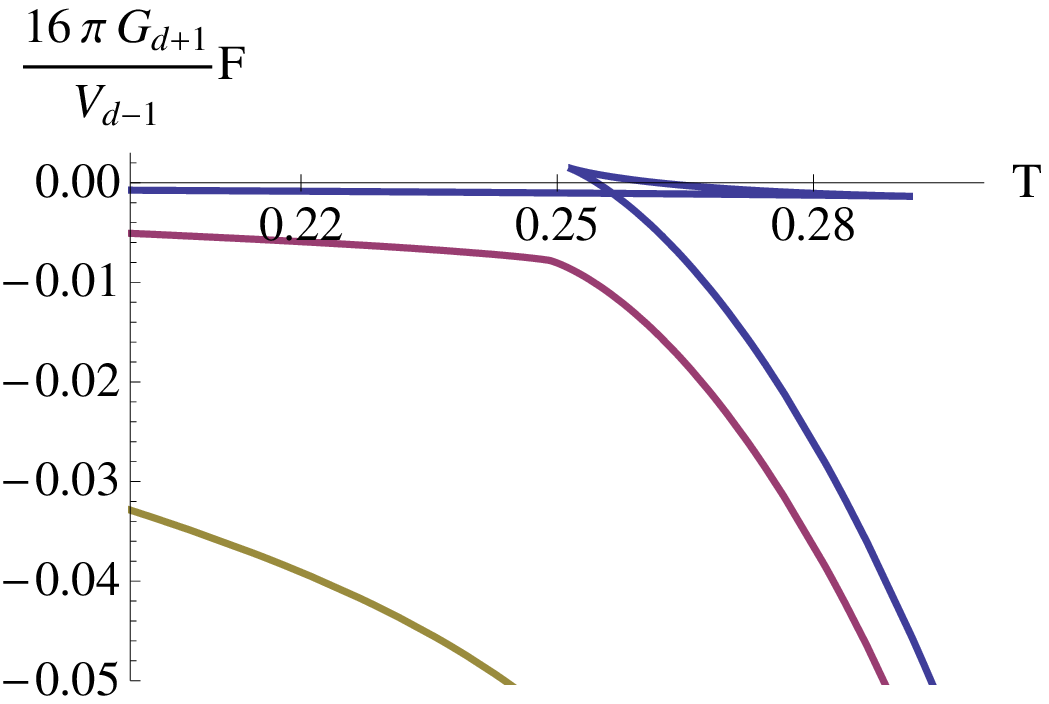}}
\caption{(Left) Temperature vs. horizon radius for $\hq=0.2 \hq_{crit}$ (blue), $\hq= \hq_{crit}$ (red),  and $\hq=5 \hq_{crit}$ (yellow). The parameters used are $d=4$, $z=1.6$, $\mu=2$ and $\ell=1$. The position of the critical radius is marked with the dashed black line. (Right) Same color code for the free energy as a function of the temperature. The branch with $\partial T/ \partial r_h<0$ in the left hand side plot corresponds to the unstable phase given by the cusp in the free energy plot. When $T\to 0$ the free energy approaches $0$ from below.}\label{fig.betafixedcharge}
\end{center}
\end{figure}

Analyzing the temperature from \eqref{eq.temperature} for fixed charge, we observe that it can present an inflexion point when plotted against $r_h$. This happens at a position $r_{crit}$ when the charge has the specific value $\hq_{crit}$
\be\label{eq.criticals}
r_{crit}^2  =  \frac{(2-z)(d-2)^2 \ell^2}{z(d+z-2)(d+z-1)} \ , \quad
\hq_{crit}^2  =  \frac{2z(d-1)(d+z-1)\mu^{\sqrt{2\frac{z-1}{d-1}}} \ell^{-2z}}{(d+z-3)(2d+z-4)}r_{crit}^{2(d+z-2)}\ .
 \ee
 For values of the charge $\hq<\hq_{crit}$ there is a region of temperatures described by three different black holes, with different horizon radii. If the charge is larger than the critical value then the relation between temperature and black-hole radius is in one-to-one correspondence. This situation is analogous to the one encountered for $z=1$ \cite{Chamblin:1999tk}, and can be seen in figure \ref{fig.betafixedcharge}. In that figure we present also the results for the free energy, which show that the branch with $\partial T/ \partial r_h<0$, for charges below the critical one, is unstable, and a first order phase transition is present for these values of the physical parameters. At $\hq=\hq_{crit}$ the kink in the free energy disappears and for larger values of the charge there is no phase transition.

\begin{figure}[tb]
\begin{center}
\subfigure{\includegraphics[scale=0.625]{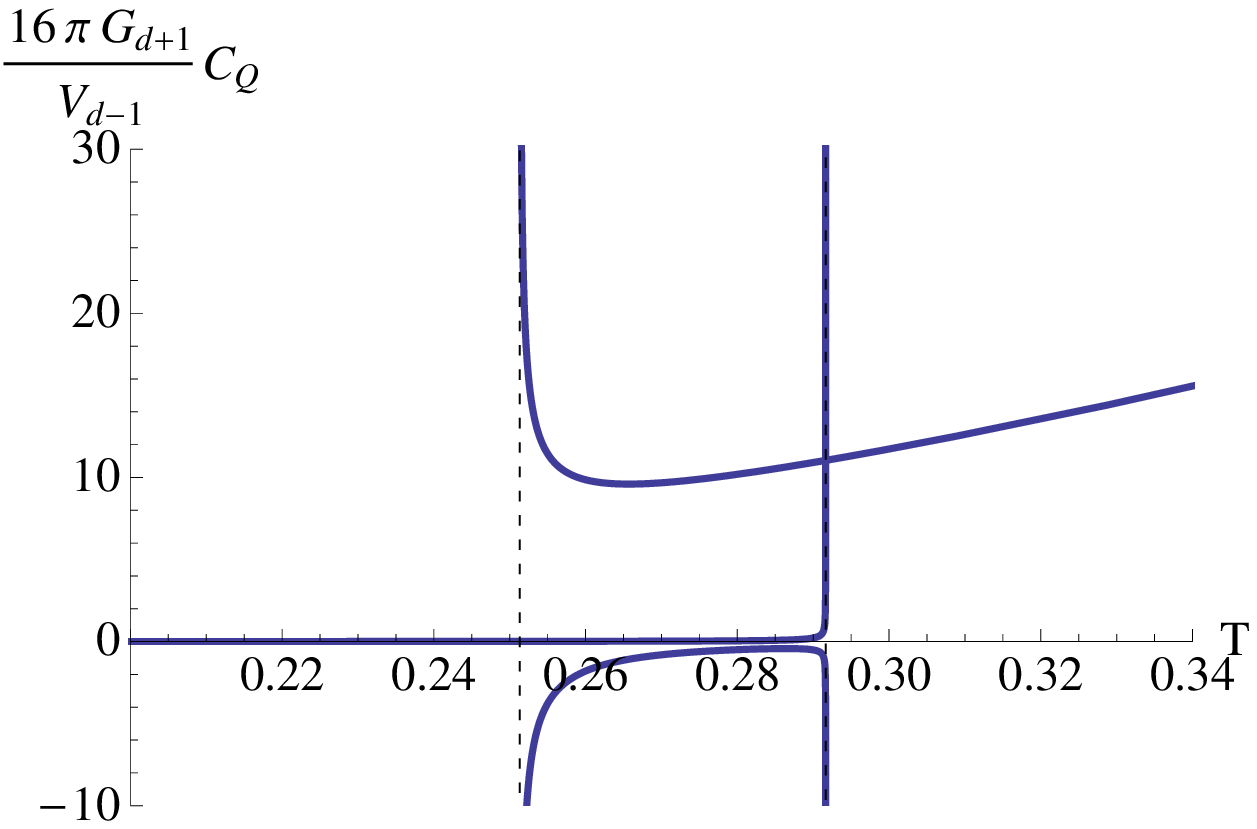}}
\subfigure{\includegraphics[scale=0.625]{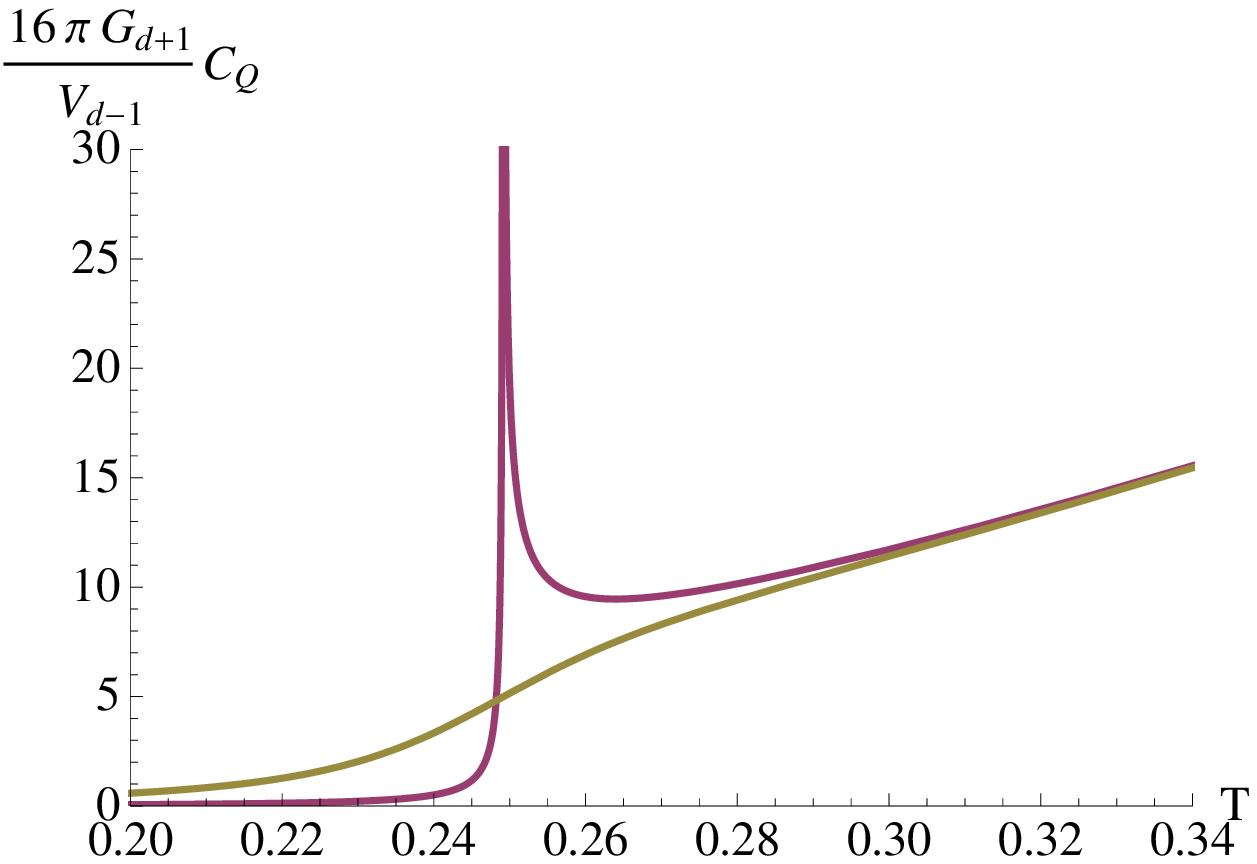}}
\caption{Heat capacity for black holes with charge less than, equal to and greater than the critical charge. The coloring is the same as in figure \ref{fig.betafixedcharge}. We have split the plot in two graphs for clarity.}\label{fig.heatcap}
\end{center}
\end{figure}

Another way to study this case is to consider the heat capacity at constant charge derived in \eqref{eq.heatcap}. The heat capacity associated to the unstable branch turns out to be negative, whereas the heat capacity for the two other branches, even beyond the kink in the free energy, is positive. In fact, one could overheat or undercool the system, keeping it in a metastable phase. In this case, the heat capacity grows as we enter further into the metastable region, eventually diverging. Figure \ref{fig.heatcap} shows this behaviour.

\paragraph{The $z=2$ case}

When $z=2$ something peculiar happens. Looking at the expressions for the critical radius and charge \eqref{eq.criticals}, we see that it occurs at the origin of spacetime, and that the critical value corresponds to the uncharged black hole. Any charge will be greater than $\hq_{crit}$ and we do not expect a phase transition. This can be seen by noticing that the term taking into account finite size effects in \eqref{eq.canfreeen} has disappeared. Therefore, for this particular value of the dynamic exponent, the free energy expression is equal to the one corresponding to the black brane solution\footnote{Except for the factor $V_{d-1}$, which in this black hole case is finite, whereas in the black brane setup is infinity.}. This means that, effectively, there are not two scales to compare (namely the horizon size and radius of the $S^{d-1}$) and therefore no phase transitions are present (notice however that the expression for the temperature still feels the finite size effects, so this is not quite the same as the  study of the planar black brane). The heat capacity is positive for every value of the charge, signal of the absence of thermodynamic instabilities in the system for this value of the dynamical exponent $z$.

\paragraph{The $z>2$ case}

For $z>2$ the term in the free energy proportional to $k(z-2)$ has reversed sign. One can show that the temperature is a bijective function of  the horizon radius, which has a minimal value at which the black hole becomes extremal (the situation is completely analogous to the one shown in figure \ref{fig.largephi}). The free energy is negative everywhere except at $T=0$ where it vanishes, and there is no non-trivial phase structure. Equivalently, the heat capacity at fixed charge is always positive.

\section{Final comments} \label{sec.remarks}

We have determined that for asymptotically Lifshitz spacetimes with a charged black hole in the center of the geometry, there is a phase structure that depends crucially on the dynamical exponent $z$.

For values $1\leq z \leq 2$ the situation is reminiscent of that of the asymptotically AdS case, which corresponds to setting $z=1$. In the grand canonical ensemble there is a line of first order phase transitions. For large values of the temperature or the potential the thermodynamics are determined by the black hole configuration, whereas for low values of both parameters the ensemble is dominated by Lifshitz spacetime. When the temperature vanishes, for values of the potential larger than the critical one given in \eqref{eq.criticalpotential}, the description is given in terms of extremal black holes with a nonzero entropy density.

\begin{figure}[tb]
\begin{center}
\subfigure{\includegraphics[scale=0.7]{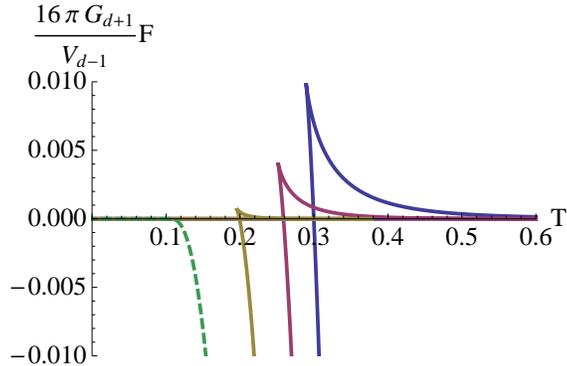}}
\caption{Free energy for the zero charge case with parameters $d=4$, $\mu=2$, $\ell=1$ and $z=1.5$ (blue), $z=1.6$ (red), $z=1.75$ (yellow), and $z=2$ (dashed green). The kink is situated at $T=T_{min}$ and the curve with large gradient touches the $F=0$ axis at $T=T_{HP}$. For larger values of $z$ the free energy is negative for all values of the temperature except $T=0$, where it vanishes. There is a branch with zero free energy from $T=0$ up to a finite value of the temperature. Then there is a kink at $F=0$ which gives rise to a branch with positive values of the free energy, until $T_{min}$ is reached. The right-most kink (not showed) gets closer to $T_{HP}$ as $z$ is increased, until it coincides when $z=2$.}\label{fig.zerocharge}
\end{center}
\end{figure}

 In the canonical ensemble there is also a line of first order phase transitions between two black hole setups, ending at a critical point given by the coordinates
 \beqa
T_{crit} & = &  \frac{(d-2)^z  }{\pi  \,z (2 d+z-4) \ell} \left[\frac{2-z}{z (d+z-2) (d+z-1)}\right]^{\frac{z-2}{2}} \\
\hq_{crit}^2 & = &   \frac{2z(d-1)(d+z-1)\mu^{\sqrt{2\frac{z-1}{d-1}}}}{(d+z-3)(2d+z-4) \ell^{2z} } \left[  \frac{(2-z)(d-2)^2 \ell^{2}}{z(d+z-2)(d+z-1)} \right]^{d+z-2}  \ .
 \eeqa
 For values of the charge greater than $\hq_{crit}$ there is no phase transition. For $\hq=0$ a Hawking-Page transition exists, with the thermodynamic ensemble dominated by Lifshitz spacetime at low temperatures and by the black hole solution for large temperatures. The exact point at which this phase transition takes place can be determined by calculating the non-trivial  radius of the horizon at which the uncharged solution has vanishing free energy, and then plugging in equation \eqref{eq.temperature}, obtaining
 \be
 T_{HP} =   \frac{d-1}{2\pi (2-z) \ell} \left[ \frac{(2-z)(d-2)^2}{z(d+z-3)^2} \right]^{z/2} \ .
 \ee
Notice that the position of the critical point in the $\hq-T$ plane depends crucially on $z$. Indeed, for the special case $z=2$ one cannot find signs of a phase transition of any order at a finite value of the charge, and only the Hawking-Page transition remains at a temperature given by $T_{HP,z=2} = (d-2)^2/4\pi(d-1) \ell$. At precisely this value of the dynamic index, the temperatures $T_{min}$ and $T_{HP}$ coincide. To understand why, it is useful to plot the free energy for zero charge and various values of $z$. We do this in figure \ref{fig.zerocharge}. There we see that the value of the temperature at which the free energy presents a kink and is positive is given by $T_{min}$, whereas this non-trivial branch crosses the horizontal axis at $T_{HP}$. When we increase the value of the dynamical exponent $z$, the kink gets closer to the $T$-axis, and at precisely $z=2$ it sits on top of it. If one increases further the value of the dynamical exponent, the free energy is negative for any value of the temperature, and therefore the Hawking-Page transition disappears, the uncharged solution being described by a black hole except at $T=0$.

For $z>2$ there is a dramatic change in the phase diagrams, the phase transition disappears in both the canonical and grand-canonical ensembles, and the black hole description is the dominant one for any value of $ \Phi,T$ (or $\hq,T$) except at the points $( \Phi_c,T=0)$, as given by equation \eqref{eq.grandcanT0}, and  $(\hq=0,T=0)$.

We have shown that the instabilities associated to thermal fluctuations, given by black holes with a negative heat capacity as given by equation \eqref{eq.heatcap}, correspond to thermodynamically unfavored phases. However, one should worry about electric instabilities as well. The isothermal susceptibility is given by
\be
\chi \equiv \left( \frac{\partial  Q}{\partial  \Phi} \right)_T \ ,
\ee
and has to be non-negative for the configuration to be stable. For $1\leq z<2$, as was the case in \cite{Chamblin:1999hg}, there is a region in the $( Q,T)$ plane where such an electric instability is present. The line $ Q_{ins}(T)$ splitting the phase space into stable and unstable regions is given by the condition $\chi( Q_{ins},T) = \infty $, leading to`
\be
 \hq_{ins}^2 = 2(d-1) \mu^{\sqrt{2\frac{z-1}{d-1}}} \ell^{2(d-2)} \left( \frac{2\pi T\ell (2-z)}{d+z-1} \right)^{2\frac{d+z-3}{z}} \left[ \frac{(d-2)^2}{d+z-3} - \frac{z(d+z-1)}{2-z} \left( \frac{2\pi T\ell (2-z)}{d+z-1} \right)^{\frac{2}{z}} \right] \ .
\ee
This curve exists only for $1\leq z<2$, and encloses a region (for a given temperature, the values of the charge lower than $ Q_{ins}(T)$) in which the solution presented here is unstable under electric perturbations. The critical point at which the line of first order phase transitions finishes is unstable. However, the temperature at which the Hawking-Page transition occurs is in the stable region, approaching it as $z\to2$.

\subsection*{Further directions}

We end with some comments for further study. As mentioned in the introduction, the Lifshitz background supported by a dilatonic scalar suffers from divergencies at the boundary, which complicates a proper holographic formulation.
Other issues related to Lifshitz spacetimes were discussed for example in \cite{Hoyos:2010at,Copsey:2010ya}. It is expected that some of these problems will be resolved by studying more general models, for instance gauged supergravities with non-trivial scalar potentials that arise from string compactifications. Finding an embedding of our model into string theory is therefore worth investigating.

The phases described in this paper should have an interpretation in terms of the dual  boundary field theory. It would be interesting to identify and analyze the properties and phases of this field theory. For this, one needs to find an order parameter, which acquires different expectation values in the different phases. In the relativistic case, i.e. for $z=1$, such an order parameter is given by the Wilson loop.

It is also interesting to perform a thorough investigation of the  critical exponents at the phase transition and thermodynamic instabilities, as was done in the AdS case in  \cite{Niu:2011tb}.The critical exponents may be sensitive to the value of the dynamical exponent $z$, as might the properties of the universality classes. 
Another line of research goes along the work performed in \cite{Pang:2009wa}, where transport coefficients were studied in asymptotically Lifshitz spacetimes.

Finally, another extension of our model is to add bulk fermions in the probe approximation. This allows us to study properties of condensed matter systems with fermions that obey Lifshitz scaling. A first step in this direction is under investigation \cite{GPSV}.

\section*{Acknowledgments}

J.T. is thankful to the Front of Galician-speaking Scientists for encouragement. J.T. is supported by the Netherlands Organization for Scientic Research (NWO) under the FOM Foundation research program. We further acknowledge support by the Netherlands Organization for Scientific Research (NWO) under the VICI grant 680-47-603.


\begin{thebibliography}{99}

\bibitem{Maldacena:1997re}
  J.~M.~Maldacena,
  ``The Large N limit of superconformal field theories and supergravity,''
  Adv.\ Theor.\ Math.\ Phys.\  {\bf 2 } (1998)  231-252.
  [hep-th/9711200].
  
\bibitem{Son:2008ye}
  D.~T.~Son,
  ``Toward an AdS/cold atoms correspondence: A Geometric realization of the
  Schrodinger symmetry,''
  Phys.\ Rev.\  D {\bf 78}, 046003 (2008)
  [arXiv:0804.3972 [hep-th]].
  
\bibitem{Balasubramanian:2008dm}
  K.~Balasubramanian and J.~McGreevy,
  ``Gravity duals for non-relativistic CFTs,''
  Phys.\ Rev.\ Lett.\  {\bf 101}, 061601 (2008)
  [arXiv:0804.4053 [hep-th]].
  

\bibitem{Kachru:2008yh}
  S.~Kachru, X.~Liu, M.~Mulligan,
  ``Gravity Duals of Lifshitz-like Fixed Points,''
  Phys.\ Rev.\  {\bf D78 } (2008)  106005.
  [arXiv:0808.1725 [hep-th]].

\bibitem{Maldacena:2008wh}
  J.~Maldacena, D.~Martelli and Y.~Tachikawa,
  ``Comments on string theory backgrounds with non-relativistic conformal
  symmetry,''
  JHEP {\bf 0810} (2008) 072.
  [arXiv:0807.1100 [hep-th]].
  
\bibitem{Ross:2009ar}
  S.~F.~Ross, O.~Saremi,
  ``Holographic stress tensor for non-relativistic theories,''
  JHEP {\bf 0909 } (2009)  009.
  [arXiv:0907.1846 [hep-th]].
  
\bibitem{Balasubramanian:2010uw}
  K.~Balasubramanian and J.~McGreevy,
  ``The Particle number in Galilean holography,''
  JHEP {\bf 1101} (2011) 137
  [arXiv:1007.2184 [hep-th]].

\bibitem{Guica:2010sw}
  M.~Guica, K.~Skenderis, M.~Taylor and B.~C.~van Rees,
  ``Holography for Schrodinger backgrounds,''
  JHEP {\bf 1102} (2011) 056.
  [arXiv:1008.1991 [hep-th]].
  
\bibitem{Brynjolfsson:2009ct}
  E.~J.~Brynjolfsson, U.~H.~Danielsson, L.~Thorlacius and T.~Zingg,
  ``Holographic Superconductors with Lifshitz Scaling,''
  J.\ Phys.\ A  {\bf 43} (2010) 065401
  [arXiv:0908.2611 [hep-th]].
  
\bibitem{Balasubramanian:2009rx}
  K.~Balasubramanian, J.~McGreevy,
  ``An Analytic Lifshitz black hole,''
  Phys.\ Rev.\  {\bf D80 } (2009)  104039.
  [arXiv:0909.0263 [hep-th]].
  
\bibitem{AyonBeato:2009nh}
  E.~Ayon-Beato, A.~Garbarz, G.~Giribet, M.~Hassaine,
  ``Lifshitz Black Hole in Three Dimensions,''
  Phys.\ Rev.\  {\bf D80 } (2009)  104029.
  [arXiv:0909.1347 [hep-th]].
  
\bibitem{Cai:2009ac}
  R.~G.~Cai, Y.~Liu, Y.~W.~Sun,
  ``A Lifshitz Black Hole in Four Dimensional R**2 Gravity,''
  JHEP {\bf 0910 } (2009)  080.
  [arXiv:0909.2807 [hep-th]].
  
\bibitem{Pang:2009pd}
  D.~W.~Pang,
  ``On Charged Lifshitz Black Holes,''
  JHEP {\bf 1001 } (2010)  116.
  [arXiv:0911.2777 [hep-th]].
  
\bibitem{AyonBeato:2010tm}
  E.~Ayon-Beato, A.~Garbarz, G.~Giribet, M.~Hassaine,
  ``Analytic Lifshitz black holes in higher dimensions,''
  JHEP {\bf 1004 } (2010)  030.
  [arXiv:1001.2361 [hep-th]].
  
\bibitem{Dehghani:2011tx}
  M.~H.~Dehghani, R.~B.~Mann and R.~Pourhasan,
  ``Charged Lifshitz Black Holes,''
  arXiv:1102.0578 [hep-th].

\bibitem{Chemissany:2011mb}
  W.~Chemissany and J.~Hartong,
  ``From D3-Branes to Lifshitz Space-Times,''
  arXiv:1105.0612 [hep-th].
  
\bibitem{Maeda:2011jj}
  H.~Maeda and G.~Giribet,
  ``Lifshitz black holes in Brans-Dicke theory,''
  arXiv:1105.1331 [gr-qc].
  
\bibitem{Danielsson:2009gi}
  U.~H.~Danielsson, L.~Thorlacius,
  ``Black holes in asymptotically Lifshitz spacetime,''
  JHEP {\bf 0903 } (2009)  070.
  [arXiv:0812.5088 [hep-th]].
  
\bibitem{Mann:2009yx}
  R.~B.~Mann,
  ``Lifshitz Topological Black Holes,''
  JHEP {\bf 0906 } (2009)  075.
  [arXiv:0905.1136 [hep-th]].
  
\bibitem{Bertoldi:2009vn}
  G.~Bertoldi, B.~A.~Burrington, A.~Peet,
  ``Black Holes in asymptotically Lifshitz spacetimes with arbitrary critical exponent,''
  Phys.\ Rev.\  {\bf D80 } (2009)  126003.
  [arXiv:0905.3183 [hep-th]].
  
\bibitem{Bertoldi:2009dt}
  G.~Bertoldi, B.~A.~Burrington, A.~W.~Peet,
  ``Thermodynamics of black branes in asymptotically Lifshitz spacetimes,''
  Phys.\ Rev.\  {\bf D80 } (2009)  126004.
  [arXiv:0907.4755 [hep-th]].
  
  
\bibitem{Dehghani:2010kd}
  M.~H.~Dehghani, R.~B.~Mann,
  ``Lovelock-Lifshitz Black Holes,''
  JHEP {\bf 1007 } (2010)  019.
  [arXiv:1004.4397 [hep-th]].
  
\bibitem{Brenna:2011gp}
  W.~G.~Brenna, M.~H.~Dehghani, R.~B.~Mann,
  ``Quasi-Topological Lifshitz Black Holes,''
  [arXiv:1101.3476 [hep-th]].
  
\bibitem{Amado:2011nd}
  I.~Amado and A.~F.~Faedo,
  ``Lifshitz black holes in string theory,''
  arXiv:1105.4862 [hep-th].
  
\bibitem{Taylor:2008tg}
  M.~Taylor,
  ``Non-relativistic holography,''
  [arXiv:0812.0530 [hep-th]].
  
\bibitem{Balasubramanian:2010uk}
  K.~Balasubramanian, K.~Narayan,
  ``Lifshitz spacetimes from AdS null and cosmological solutions,''
  JHEP {\bf 1008}, 014 (2010).
  [arXiv:1005.3291 [hep-th]].
  
\bibitem{Donos:2010tu}
  A.~Donos and J.~P.~Gauntlett,
  ``Lifshitz Solutions of D=10 and D=11 supergravity,''
  JHEP {\bf 1012} (2010) 002
  [arXiv:1008.2062 [hep-th]].
  
\bibitem{Cassani:2011sv}
  D.~Cassani, A.~F.~Faedo,
  ``Constructing Lifshitz solutions from AdS,''
  JHEP {\bf 1105}, 013 (2011).
  [arXiv:1102.5344 [hep-th]].
    
\bibitem{Halmagyi:2011xh}
  N.~Halmagyi, M.~Petrini and A.~Zaffaroni,
  ``Non-Relativistic Solutions of N=2 Gauged Supergravity,''
  arXiv:1102.5740 [hep-th].
  
\bibitem{Goldstein:2009cv}
  K.~Goldstein, S.~Kachru, S.~Prakash and S.~P.~Trivedi,
  ``Holography of Charged Dilaton Black Holes,''
  JHEP {\bf 1008} (2010) 078
  [arXiv:0911.3586 [hep-th]].
   
\bibitem{Charmousis:2010zz}
  C.~Charmousis, B.~Gouteraux, B.~S.~Kim, E.~Kiritsis and R.~Meyer,
  ``Effective Holographic Theories for low-temperature condensed matter
  systems,''
  JHEP {\bf 1011} (2010) 151
  [arXiv:1005.4690 [hep-th]].
     
     
\bibitem{Hawking:1982dh}
  S.~W.~Hawking, D.~N.~Page,
  ``Thermodynamics of Black Holes in anti-De Sitter Space,''
  Commun.\ Math.\ Phys.\  {\bf 87 } (1983)  577.
  
\bibitem{Witten:1998zw}
  E.~Witten,
  ``Anti-de Sitter space, thermal phase transition, and confinement in gauge theories,''
  Adv.\ Theor.\ Math.\ Phys.\  {\bf 2 } (1998)  505-532.
  [hep-th/9803131].
  
  
\bibitem{Chamblin:1999tk}
  A.~Chamblin, R.~Emparan, C.~V.~Johnson, R.~C.~Myers,
  ``Charged AdS black holes and catastrophic holography,''
  Phys.\ Rev.\  {\bf D60 } (1999)  064018.
  [hep-th/9902170].
  
  
\bibitem{Chamblin:1999hg}
  A.~Chamblin, R.~Emparan, C.~V.~Johnson, R.~C.~Myers,
  ``Holography, thermodynamics and fluctuations of charged AdS black holes,''
  Phys.\ Rev.\  {\bf D60 } (1999)  104026.
  [hep-th/9904197].
  
  
\bibitem{Hoyos:2010at}
  C.~Hoyos, P.~Koroteev,
  ``On the Null Energy Condition and Causality in Lifshitz Holography,''
  Phys.\ Rev.\  {\bf D82 } (2010)  084002.
  [arXiv:1007.1428 [hep-th]].

  
\bibitem{Copsey:2010ya}
  K.~Copsey, R.~Mann,
  ``Pathologies in Asymptotically Lifshitz Spacetimes,''
  JHEP {\bf 1103 } (2011)  039.
  [arXiv:1011.3502 [hep-th]].
  
\bibitem{Niu:2011tb}
  C.~Niu, Y.~Tian, X.~Wu,
  ``Critical Phenomena and Thermodynamic Geometry of RN-AdS Black Holes,''
  [arXiv:1104.3066 [hep-th]].
  
\bibitem{Pang:2009wa}
  D.~W.~Pang,
  ``Conductivity and Diffusion Constant in Lifshitz Backgrounds,''
  JHEP {\bf 1001 } (2010)  120.
  [arXiv:0912.2403 [hep-th]].
  
  \bibitem{GPSV}
  U. G\"ursoy, E. Plauschinn, H. Stoof, S. Vandoren, work in progress.

\end{thebibliography}
\end{document}